\documentclass[jpsj_ref]{jpsj2}
\usepackage{graphicx}
\usepackage{bm}
\newcommand{\half}{\frac{1}{2}} 
\def\average#1{\left\langle {#1} \right\rangle}

\def\deld#1#2{\frac{\delta #1}{\delta #2}}
\def\vvec#1{\stackrel{{\leftrightarrow}}{#1}} 
\def\vec2#1#2{\left(\begin{array}{c} #1 \\ #2 \end{array}\right)}
\def\vec3#1#2#3{\left(\begin{array}{c} #1 \\ #2 \\ #3 \end{array}\right)}

\newcommand{\adag}{{a^{\dagger}}}

\newcommand{\Av}{{\bm A}}

\newcommand{\Bv}{{\bm B}}

\newcommand{\cdag}{{c^{\dagger}}}

\newcommand{\ckvs}{c_{\kv\sigma}}

\newcommand{\dx}{{d^3 x}}
\newcommand{\DOS}{{\nu}}
\newcommand{\DOSV}{{N(0)}}

\newcommand{\eF}{{\epsilon_F}}

\newcommand{\ekvs}{\epsilon_{\kv\sigma}}
\newcommand{\Ev}{{\bm E}}

\newcommand{\evth}{{\bm e}_{\theta}}
\newcommand{\evph}{{\bm e}_{\phi}}
\newcommand{\evs}{{\bm n}}

\newcommand{\evsph}{{\bm e}_{{\rm s}\phi}}
\newcommand{\evz}{{\bm e}_{z}}

\newcommand{\Fena}{{F_{\rm e}^{\rm na}}}
\newcommand{\Fv}{{\bm F}}
\newcommand{\gv}{{\bm g}}
\newcommand{\gr}{g^{\rm r}}
\newcommand{\ga}{g^{\rm a}}

\newcommand{\hf}{\frac{1}{2}}

\newcommand{\He}{H_{\rm e}}

\newcommand{\Hem}{H_{\rm EM}}
\newcommand{\Hex}{H_{\rm ex}}
\newcommand{\Himp}{H_{\rm imp}}

\renewcommand{\Im}{{\rm Im}}

\newcommand{\intx}{\int {d^3x}}
\newcommand{\iv}{\bm{i}}

\newcommand{\js}{j_{\rm s}}

\newcommand{\jsv}{\bm{j}_{\rm s}}

\newcommand{\jv}{\bm{j}}

\newcommand{\kb}{{k_B}}
\newcommand{\kv}{{\bm k}}

\newcommand{\kf}{{k_F}}

\newcommand{\Kp}{{K_\perp}}
\newcommand{\ktil}{\tilde{k}}

\newcommand{\lamv}{{\lambda_{\rm v}}}

\newcommand{\Le}{{L_{\rm e}}}

\newcommand{\mv}{{\bm m}}

\newcommand{\muB}{\mu_B}
\newcommand{\Ne}{N_{\rm e}}

\newcommand{\nvortex}{n_{\rm v}}

\newcommand{\Omz}{{{\Omega_0}}}

\newcommand{\pv}{{\bm p}}
\newcommand{\qv}{{\bm q}}
\newcommand{\qtil}{{\tilde{q}}}

\newcommand{\rhos}{{\rho_{\rm S}}}
\newcommand{\rhoS}{{\rho_{\rm S}}}
\newcommand{\rhoxy}{{\rho_{xy}}}
\newcommand{\RS}{{R_{\rm S}}}

\newcommand{\Rv}{{\bm R}}
\newcommand{\sigmav}{{\bm \sigma}}
\newcommand{\se}{{s}}
\newcommand{\sev}{{\bm \se}}

\newcommand{\svtil}{\tilde{\bm s}}
\newcommand{\stil}{\tilde{s}}
\newcommand{\stilz}{\stil_{z}}
\newcommand{\stilpara}{\stil_{\parallel}}
\newcommand{\stilperp}{\stil_{\perp}}

\newcommand{\Sv}{{{\bm S}}}

\newcommand{\sumx}{{\int \frac{d^3x}{a^3}}}

\newcommand{\sumkv}{{\sum_{\kv}}}
\newcommand{\sumom}{\int\frac{d\omega}{2\pi}}

\newcommand{\sumqv}{{\sum_{\qv}}}
\newcommand{\thickness}{{d}}

\newcommand{\torque}{{\tau}}
\newcommand{\torquev}{{\bm \torque}}

\newcommand{\thetast}{\theta_{\rm st}}

\newcommand{\Vinv}{\frac{1}{V}}

\newcommand{\xv}{{\bm x}}
\newcommand{\Xv}{{\bm X}}

\newcommand{\xw}{{z}}

\title{
Spin torque and force due to current for general spin textures}
\author{
Gen Tatara$^{1,2}$,
Hiroshi Kohno$^{3}$,
Junya Shibata$^{4}$,
Yann Lemaho$^5$,
and
Kyung-Jin Lee$^6$
}

\inst{
$^1$Graduate School of Science, Tokyo Metropolitan University\\
1-1 Minamiosawa, Hachioji,  Tokyo 192-0397\\
$^2$PRESTO, JST, 4-1-8 Honcho, Kawaguchi, Saitama 332-0012\\
$^3$
Graduate School of Engineering Science, Osaka University,
Toyonaka, Osaka 560-8531\\
$^4$
Frontier Research System (FRS), The Institute of Physical and
Chemical Research (RIKEN), 2-1 Hirosawa, Wako, Saitama 351-0198\\
$^5$
Institut d¡ÇElectronique Fondamentale, UMR CNRS 8622, University Paris Sud, Orsay, France\\
$^6$
Department of Materials Science \& Engineering,
Korea University
Anam-dong 5, Sungbuk-gu
Seoul, 136-713
Korea
}

\abst{
The nonadiabatic correction to spin transfer torque arising from fast-varying spin texture is calculated treating the conductions 
electron fully quantum mechanically.
The torque is nonlocal in space, and is shown to be equivalent to a force (due to momentum transfer) acting on the center of mass of the texture. 
Another kind of force exists in the adiabatic regime, 
and it is identified to be of topological origin.
These forces are shown to be the counter-reactions of electric transport properties, resistivity and the Hall effect.
}

\kword{
spin torque, domain wall, magnetoresistance, threshold current, Berry phase, MRAM
}

\begin{document}
\maketitle
\section{Introduction}

Recently, torques on local spin (magnetization) induced by spin-polarized electric current and spin current are of special interest from the 
viewpoint of magnetization switching by current in wires\cite{Berger78,Berger84,Berger92} and  nanopillars\cite{Slonczewski96,Berger96}.
Berger\cite{Berger84,Berger96} and Slonczewski\cite{Slonczewski96} pointed out that the exchange of the spin angular momentum between the local spin and current-carrying conduction electron induces a torque called spin-transfer torque, and that this is a dominant mechanism of magnetization flip when the magnetization is slowy varying (i.e. in the adiabatic regime).
In the case of a domain wall, another effect arising from the current, that is, 
the effect due to elecron reflection  or  nonadiabaticity, 
has been discussed by Berger\cite{Berger78}.
This force was later derived microscopically by estimating the nonadiabatic correction of the torque due to exchange coupling\cite{TK04}. 
Recently, a new torque term in the Landau-Lifshitz(-Gilbert) equation
was proposed, which was shown to arise from spin relaxation of the conduction electron
\cite{Zhang04,Thiaville05}.
This term, sometimes called the $\beta$-term, was later derived microscopically
\cite{Tserkovnyak06,KTS06,Duine07} and also from other mechanisms\cite{Barnes05}.
This term is also sometimes called nonadiabatic torque, in that the term arises from the deviation of the electron spin from perfect adiabaticity as a result of spin relaxation.
(This effect of spin relaxation is not the nonadiabaticity in the strict sense, so
we will call it "nonadiabaticity due to spin relaxation" to avoid confusion.)

Quite recently,Waintal and Viret\cite{Waintal04} and  Stiles and coworkers\cite{Xiao06,Stiles06} studied the spatial distribution of the current-induced torque around a domain wall by solving the 
Schr\"odinger equation and found a nonlocal oscillation torque. 
This torque is due to the nonadiabaticity arising from the finite domain wall width, or in other words,
from the fast-varying component of spin texture. 
The oscillation is of period $\sim \kf^{-1}$ ($\kf$ is the Fermi wavelength) and is of quantum origin similar to Ruderman-Kittel-Kasuya-Yosida(RKKY) oscillation.
The oscillating torque is due to oscillating spin accumulation of the electrons as argued in ref. \cite{Simanek05}.
Quite recently, nonlocal oscillating torque around domain wall was numerically studied taking account of strong spin-orbit interaction on the basis of the Kohn-Luttinger model  (i.e., in magnetic semiconductors)\cite{Nguyen06}.
It was shown that the oscillating torque is asymmetric around the domain wall and that this feature results in high wall velocity.

These results can be summarized as a total torque acting on local spin, $\Sv$, given by
\begin{equation}
\torquev=-\frac{a^3}{2eS}(\jsv\cdot\nabla)\Sv
- \frac{a^3 \beta}{eS} [\Sv \times (\jsv\cdot\nabla)\Sv]
+\torquev_{\rm nl}, \label{torques}
\end{equation}
where the first term represents the spin transfer torque, the second is the $\beta$ term due to spin relaxation, and the last term denotes the nonlocal torque argued in refs. \cite{Waintal04,Xiao06,Stiles06} 
(but explicit expression was not obtained there).
We note here that the non-adiabaticity has two different origins, spin relaxation and fast-varying local spin texture, and that these two lead to totally different torques ($\beta$ term and $\torquev_{\rm nl}$).
Spin transfer torque was studied further within the Boltzmann approach in ref. \cite{Piechon06}.
So far, most numerical simulations on magnetization dynamics\cite{Thiaville05} have modeled the effect of current solely by local torque terms and nonlocal torque $\torquev_{\rm nl}$ has been neglected.
Such simulations are thus valid only in the slowly varying limit.
Although the nonlocal term is believed to be small in most cases, this term could still be important in reality, since, like the $\beta$ term it plays a qualitatively different role as spin transfer torque, and  $\beta$ term is also quite small.
In particular, the nonlocal term would be important in vortex-wall structures, where the spins vary rapidly inside the core.
As we will demonstrate, two nonadiabaticities, $\beta$ and $\torquev_{\rm nl}$, both contribute to a total force acting on spin texture.
The dynamics of a rigid domain wall under the total torque given by
eq. (\ref{torques}) was studied in ref. \cite{TTKSNF06}.

One of the aims of this study is to derive an explicit expression of nonlocal torque, $\torquev_{\rm nl}$, within the lowest order correction to the slow variation, treating the conduction electron fully quantum mechanically.
As we will show, quantum treatment is essential for discussing nonadiabatic torque and force.
In our calculation here, we assume $\Delta \tau/\hbar \gg 1$, where $\Delta$ and $\tau$ are exchange splitting and  (elastic) lifetime of the conduction electron, respectively, and takes account of the nonadiabaticity represented as higher order terms in $1/(\kf \lambda)$, $\lambda$ being the scale of the spatial variation of spin texture. 
$\Delta \tau/\hbar \gg 1$ is one definition of adiabaticity\cite{Stern92,Popp03} where the gauge field can be treated perturbatively.

Second, we will show that this complicated nonlocal torque represents the effect of electron scattering by the spin texture and that the nonlocal torque is summed up simply to a total force acting on the texture.
Third, this force is shown to be exactly proportional to resistivity due to spin structure and hence to the reflection probability.
Finally, we will show that there exists another type of force which remains finite in the adiabatic limit.
This force, adiabatic force or topological force, is proportional to the topological vortex number of the texture, as was pointed out in ref. \cite{Thiele73,KTSS06}.
It is shown to be proportional to the Hall resistivity caused by spin chirality or vortex and is thus identified with a back-reaction of the Hall effect due to spin chirality or the spin Berry phase\cite{Ye99,TKawamura02,OTN04,Taniguchi04,Fabris06}.
This force acts on a vortex or vortex wall perpendicular to the current. 
The effect of this topological force is included in the conventional spin transfer torque term,
in contrast to the nonlocal torque.

This paper is an extension of the preceeding paper\cite{TK04}.
While torque and force acting on a domain wall were exclusively studied there, those for general spin structures are studied in the present paper.
Besides, in ref. \cite{TK04}, spin transfer torque was approximated by dominant contribution (i.e., in the adiabatic limit) and non-adiabaticity was included only as a force.
This inconsistency is removed in the present paper, calculating both to the same order of non-adiabaticity.
Another aim of the present paper is to calculate the effect of current as the linear response to the applied electric field.
This was not done correctly in ref. \cite{TK04}, where the effect of current was assumed to be represented by the distribution function of current-carrying state.

\section{Model and Method}
The Hamiltonian we consider is the standard exchange interaction one given by
\begin{eqnarray}
\He 
&=& \sum_{\kv \sigma} \ekvs \cdag_{\kv\sigma}\ckvs 
   + \Himp, 
\label{H0}
\end{eqnarray}
where electron operators are denoted by $\ckvs$ and $\cdag_{\kv\sigma}$, $\ekvs\equiv \frac{k^2}{2m}-\eF$,
$\sigma=\pm$ denotes spin, and $\Himp$ represents the elastic scattering by impurities (without spin flip).
Conduction electrons are coupled to local spin, $\Sv$, {\it via} 
the exchange interaction 
\begin{eqnarray}
 \Hex 
&=& - \frac{\Delta}{S} 
\intx  \Sv(\xv) \cdot \cdag_{\xv}\sigmav c_{\xv} , 
\label{Hsd}
\end{eqnarray} 
where ${\bm \sigma}$ is the Pauli spin-matrix vector, 
and $\Delta$ is the exchange coupling constant. 
The Lagrangian of the conduction electron is defined as
$\Le\equiv \hbar \intx i \cdag \dot{c} -\He$.

All the torque acting on the local spin caused by the electron is thus obtained from the exchange term as
\begin{equation}
\torque(\xv) \equiv \frac{\delta \Hex}{\delta \Sv(\xv)} \times \Sv(\xv)
= - \frac{\Delta}{S} \Sv(\xv) \times \sev(\xv),
\end{equation}
where
\begin{equation}
\sev(\xv) \equiv \average{\cdag_{\xv}\sigmav c_{\xv} }
\end{equation}
is the spin polarization of the conduction electron evaluated in the presence of spin texture, $\Sv$.
Our task is merely to carefully estimate this spin density.

To incorporate the spatially varying spin texture, we consider the slowly varying case,  
called the adiabatic case. 
In this case, gauge transformation in spin space is useful.
We consider only the lowest-order contribution from the gauge field, but will go beyond the adiabatic limit by taking account of the nonadiabaticity represented by the finite momentum transfer ($\qv$ defined below) between the electron and gauge field.
Strictly, this expansion is justified in the limit of $\Delta\tau/\hbar \gg 1$\cite{Stern92,Popp03}.

Gauge transformation, defined by a $2\times2$ matrix, $U(\xv,t)$, 
relates the electron operator $c$ to a new operator $a$ as
\begin{equation}
c(\xv,t)\equiv U(\xv,t) a(\xv,t),
\end{equation}
where $U$ is defined as
\begin{equation}
U(\xv,t)\equiv \mv(\xv,t)\cdot\sigmav,
\end{equation}
$\mv$ being a real three-component unit vector.
The matrix satisfies $U^2=1$, or $U^{-1}=U$.
The aim of our gauge transform is to let this spin to be along the $z$-axis, {\it i.e.},
$\frac{1}{S}U^\dagger (\Sv\cdot\sigmav) U=
2 \mv(\mv\cdot\sigmav)-\sigmav=\sigmav\cdot\evz$.
This is satisfied if we choose
\begin{equation}
\mv=\left(
\sin\frac{\theta}{2}\cos\phi,\sin\frac{\theta}{2}\sin\phi,\cos\frac{\theta}{2} \right).
\end{equation}
We see that a gauge field appears from the derivative terms,  as seen in
\begin{equation}
\partial_\mu c=U(\partial_\mu+U^{-1}\partial_\mu U)a
  = U(\partial_\mu+iA_\mu)a ,
\end{equation}
where a gauge field is defined as
\begin{equation}
A_\mu\equiv -iU^{-1}\partial_\mu U.
\end{equation}
In terms of $\mv$, $A_\mu$ is written as
\begin{equation}
A_\mu=(\mv\times\partial_\mu \mv)\cdot \sigmav \equiv A_\mu^\alpha \sigma_\alpha  ,
\end{equation}
where the summation over $\alpha=x,y,z$ is suppressed.

The electron part of the Lagrangian is written in terms of the $a$-electron as
\begin{eqnarray}
\Le &=&  \sumkv \left[
i\hbar \adag \dot{a} -\epsilon_{\kv\sigma} \adag_{\kv\sigma} a_{\kv\sigma}
-\hbar\sum_{\qv,\mu}
  \left(J_\mu\left(\kv+\frac{\qv}{2}\right)\cdot A_\mu^\alpha(-\qv) \right)
  \adag_{\kv+\qv}\sigma_\alpha a_{\kv} \right.  \nonumber\\ && \left.
 -
 \frac{\hbar^2}{2m}\sum_{\qv\pv}A_i^\alpha(-\qv-\pv)A_i^\alpha(\pv)
 \adag_{\kv+\qv} a_{\kv}
 \right],\label{Le}
\end{eqnarray}
where
\begin{equation}
\epsilon_{\kv\sigma} \equiv \frac{\hbar^2 k^2}{2m}-\eF-\sigma\Delta
\end{equation}
is the electron energy in a uniform field along the $z$-axis
and
\begin{equation}
J_\mu(\kv)\equiv \left( \frac{\hbar}{m}\kv, 1  \right)
\end{equation}
for $\mu=x,y,z,t$.
In this paper, we assume slowly varying $A_\mu^\alpha$  in time and neglect the $t$-dependence.

The electron spin is modified by the gauge transformation and is written using the spin density in the gauge-transformed frame, $\svtil$, as
\begin{equation}
\sev(\xv,t)\equiv \average{\cdag\sigmav c}
  =2\mv(\mv\cdot\svtil)-\svtil,
\end{equation}
where 
\begin{equation}
\svtil(\xv,t) \equiv  \average{\adag\sigmav a}.
\end{equation}
The original electron spin is written as
\begin{equation}
\sev = -\stilpara \evth +\stilz \evs -\stilperp \evph,
\label{svex}
\end{equation}
where 
$\stilpara$ and $\stilperp $ are defined as
\begin{eqnarray}
\stilpara &\equiv & \hf \sum_{\pm} e^{\mp i\phi} \stil^{\pm}
   = \cos\phi \stil_{x} + \sin\phi \stil_{y}  \nonumber \\
\stilperp &\equiv & \hf \sum_{\pm} (\mp)ie^{\mp i\phi} \stil^{\pm}
   = -\sin\phi \stil_{x} + \cos\phi \stil_{y}, \label{stildef}
\end{eqnarray}
with
$\stil^{\pm}\equiv \stil_{x}\pm i\stil_{y}$.
The three unit vectors are
$\evs\equiv
 \Sv/S=(\sin\theta\cos\phi,\sin\theta\sin\phi,\cos\theta)$, 
\begin{eqnarray}
\evth &=& \left(   \cos\theta \cos\phi ,
  \cos\theta \sin\phi ,
   -\sin\theta  \right)  \\
\evph &=& \left(
   -\sin\phi ,    \cos\phi ,      0  \right).
\end{eqnarray}

\section{Spin Density in Linear Response}
\begin{figure}[tbh]
  \begin{center}
  \includegraphics[scale=0.3]{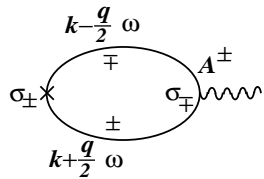}
  \end{center}
\caption{
Elecron spin density without current at the lowest order in gauge field (represented by wavy lines).
\label{FIGse0}
}
\end{figure}

We will estimate spin density $\svtil(\xv,t)$ to the lowest order in the gauge field.
Electron spin is modified by spin texture even in the absence of current.
(This effect was not considered in ref. \cite{TK04}).
This equilibrium contribution is given by (Fig. \ref{FIGse0})
\begin{eqnarray}
{\stil}^{\pm {(0)}} (\xv,t) & = &
-i \frac{\hbar^2}{V} \sumom \sum_{\kv\qv} e^{-i\qv\cdot\xv}
\sum_{\mu}
  J_\mu(\kv)A_\mu^\pm (\qv)
\nonumber\\
&&\times
[ g_{\kv-\frac{\qv}{2},\mp}^{\rm r} (\omega) 
 g_{\kv+\frac{\qv}{2},\pm}^{<} (\omega)
+ g_{\kv-\frac{\qv}{2},\mp}^{<}(\omega)
 g_{\kv+\frac{\qv}{2},\pm}^{\rm a} (\omega) ]
\nonumber\\
&=&
\frac{\hbar^2}{V} \sum_{\kv\qv} e^{-i\qv\cdot\xv} A_0^\pm (\qv)
\frac{f_{\kv+\frac{\qv}{2},\pm}-f_{\kv-\frac{\qv}{2},\mp}}
{\epsilon_{\kv+\frac{\qv}{2},\pm}-\epsilon_{\kv-\frac{\qv}{2},\mp}+\frac{i}{\tau}} ,
\label{sequil}
\end{eqnarray}
where $\gr_{\kv,\sigma}(\omega)\equiv 
\frac{1}{\omega-\epsilon_{\kv\sigma}+\frac{i}{2\tau}}$ is a free retarded Green function and $g^<$ denotes the Keldysh Green function.

Let us next calculate a current-driven part.
We estimate the response to a static applied electric field to the linear order.
The interaction between the electron and applied field is given as
\begin{equation}
\Hem = \int \dx \Av_{\rm EM} \cdot \jv,
\end{equation}
where
\begin{equation}
\Av_{\rm EM}= \frac{\Ev}{i\Omz}e^{-i\Omz t}
\end{equation}
is the electromagnetic gauge field, $\Ev$ is the applied electric field assumed to be spatially homogeneous, and $\Omz$ is the frequency of the applied field, which is chosen to be $\Omz\rightarrow0$ at the last stage of the calculation.
We note that this linear response calculation takes care of the current correctly, while treatment described in ref. \cite{TK04} lacks consisitency in that a generalized Fermi distribution function ($f_\kv$) in the presence of current flow was assumed there.
The total electric current is given in the gauge-transformed space as
\begin{eqnarray}
\jv &\equiv& -\int \dx \frac{e\hbar}{m}\frac{i}{2} 
(c^{\dagger}\stackrel{\leftrightarrow}{\nabla}_{z}c)
-\frac{e\hbar}{2m} \Av_{\rm EM} \sum_{\kv}
c^{\dagger}_{\kv}c_{\kv}
\nonumber\\&=&       
\frac{e\hbar}{m} \sum_{{\kv}} \kv a^{\dagger}_{{\kv}} 
a_{{\kv}}
 -\frac{e\hbar}{m} \sum_{\kv\qv\alpha}
A^\alpha _{\qv} a^{\dagger}_{\kv+\qv}\sigma^{\alpha}a_{\kv} 
 -\frac{e\hbar}{2m} \sum_{\kv}
\Av_{\rm EM} a^{\dagger}_{\kv}a_{\kv} 
 . \label{j}
\end{eqnarray}
\begin{figure}[tbh]
  \begin{center}
  \includegraphics[scale=0.3]{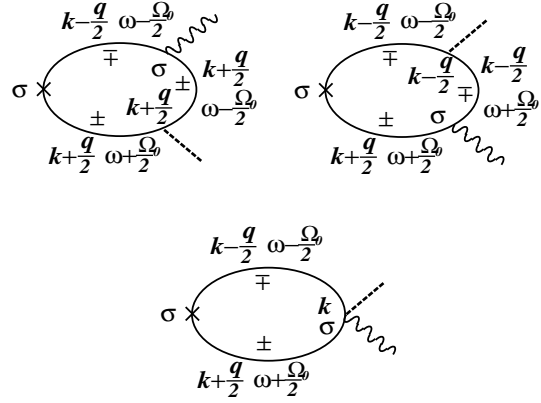}
  \end{center}
\caption{
Diagrams representing the current-driven part of the electron spin density at the linear order in both gauge field (represented by wavy lines) and  applied electric field (dotted lines).
Upper two processes are self-energy corrections and the lower process is from the correction of the current vertex.
\label{FIGsevc}
}
\end{figure}

Spin density at the first order in both of gauge fields, $\Av_{\rm EM} $ and $A^\alpha _{\qv} $, is shown in Fig. \ref{FIGsevc}.
The first two processes (denoted by (A)) are obtained as
\begin{eqnarray}
\stil^{\pm}(\xv,t)^{\rm (A)} & = &
\sum_{\omega\kv\qv} \sum_i \frac{B_i}{\Omz} 
\left[
\left( k+\frac{q}{2}\right)_i 
[g_{\kv-\frac{q}{2},\mp,\omega-\frac{\Omz}{2}} g_{\kv+\frac{q}{2},\pm,\omega-\frac{\Omz}{2}} g_{\kv+\frac{q}{2},\pm,\omega+\frac{\Omz}{2}}]^<
\right.\nonumber\\
&& \left.
+
\left(k-\frac{q}{2}\right)_i 
[g_{\kv-\frac{q}{2},\mp,\omega-\frac{\Omz}{2}} g_{\kv-\frac{q}{2},\mp,\omega+\frac{\Omz}{2}} g_{\kv+\frac{q}{2},\pm,\omega+\frac{\Omz}{2}}]^<
\right], 
\label{sSE}
\end{eqnarray}
where
\begin{equation}
B_i \equiv
 \frac{eE_i}{m V} e^{-i(\qv\cdot\xv-\Omz t)} \sum_{\mu}
 J_\mu(\kv)A_\mu^\pm (\qv) ,
\end{equation}
and the last contribution (B) arising from the modification of the 
vertex is given by
\begin{eqnarray}
\stil^{\pm}(\xv,t)^{\rm (B)} & = &
\sum_{\omega\kv\qv} \frac{B_0}{\Omz} 
\left[
g_{\kv-\frac{q}{2},\mp,\omega-\frac{\Omz}{2}}  g_{\kv+\frac{q}{2},\pm,\omega+\frac{\Omz}{2}}\right]^<,
\label{sV}
\end{eqnarray}
where
\begin{equation}
B_0\equiv
\frac{eE_i}{m V} e^{-i(\qv\cdot\xv-\Omz t)} A_i^\pm (\qv) .
\end{equation}
We will derive the leading contribution in the clean limit, $\tau\rightarrow\infty$. 
The calculation is lengthy, so we present simply the result here. 
Details are presented in the appendix. 
The result is obtained as 
($\stil^{\pm (1)}\equiv \stil^{\pm {\rm (A)}}+\stil^{\pm {\rm (B)}}$):
\begin{eqnarray}
\stil^{\pm (1)}(\xv,t)  &=&
-\frac{i}{2\pi}\sum_{\kv\qv} \sum_i  \frac{eE_i}{2 m V} e^{-i\qv\cdot\xv} \sum_{\mu}
 A_\mu^\pm (\qv) \tau
\nonumber\\ && \times 
\left[ 
 k_i  J_\mu\left(-\left(k+\frac{q}{2}\right)\right) 
 \frac{\gr_{\kv,\pm} -\ga_{\kv,\pm} }
{\epsilon_{\kv+q}-\epsilon_{\kv} \pm2\Delta} 
-  k_i J_\mu\left(\kv+\frac{q}{2}\right)
 \frac{\gr_{\kv,\mp} - \ga_{\kv,\mp}}
{\epsilon_{\kv+{q}}-\epsilon_{\kv} \mp2\Delta}
\right.\nonumber\\
&& \left.
-i\pi q_i \left(
\gr_{\kv,\mp} J_\mu\left(\kv+\frac{q}{2}\right) 
\delta({\epsilon_{\kv+q}-\epsilon_{\kv} \mp2\Delta})
-\ga_{\kv,\pm}  J_\mu\left(-\left(\kv+\frac{q}{2}\right)\right) 
\delta({\epsilon_{\kv+q}-\epsilon_{\kv} \pm2\Delta})
\right)
\right].
\nonumber\\ \label{S1}
\end{eqnarray}

\begin{figure}[tbh]
  \begin{center}
  \includegraphics[width=7cm]{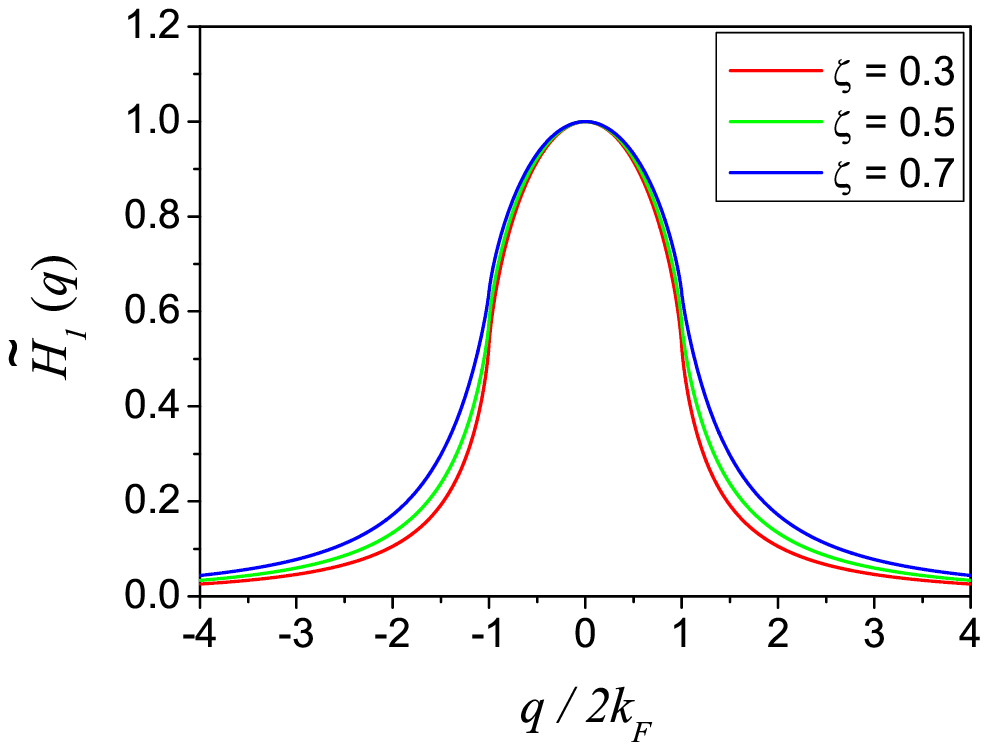}
  \includegraphics[width=7cm]{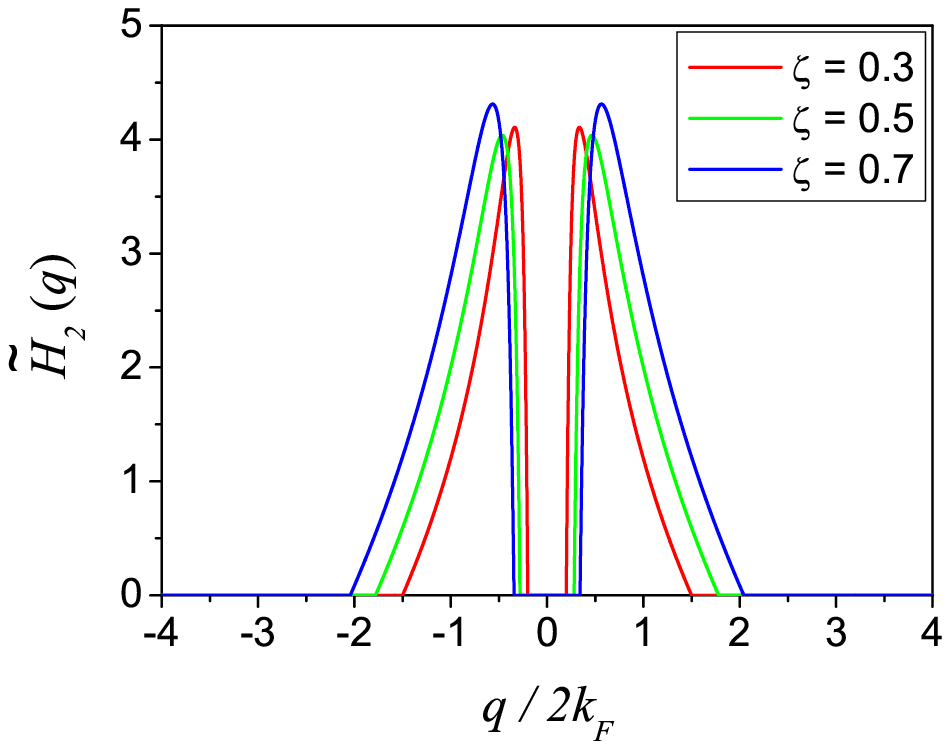}
  \includegraphics[width=7cm]{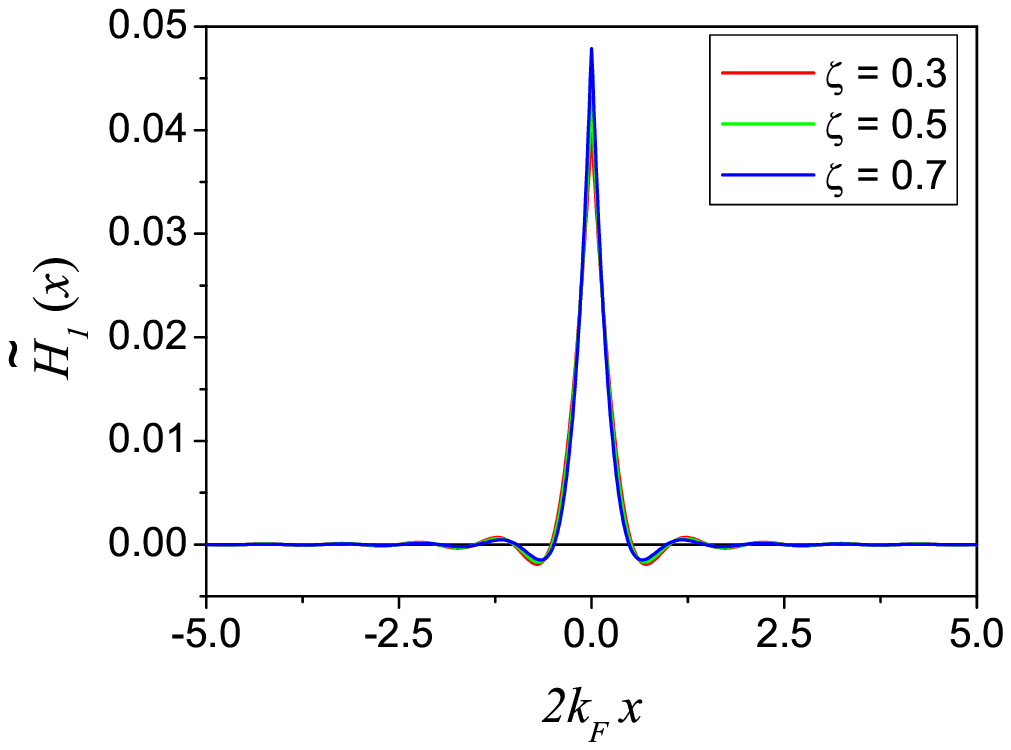}
  \includegraphics[width=7cm]{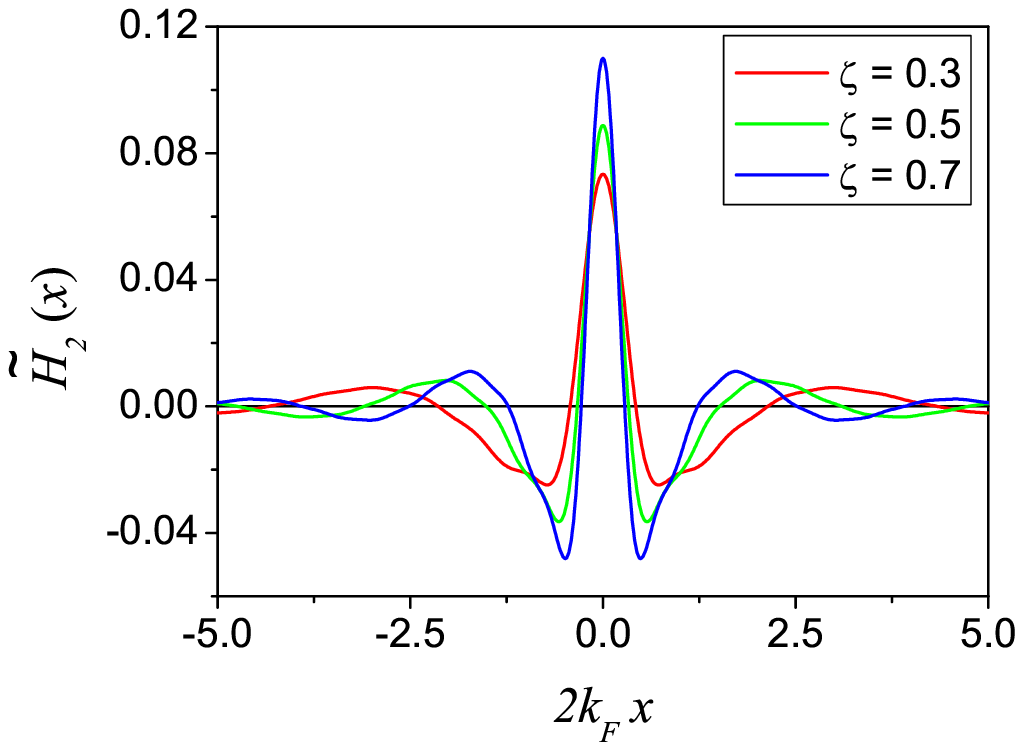}
  \end{center}
\caption{ (Color online)
Plot of correlation functions, $\tilde H_1(q)$ and $\tilde H_2(q)$, and
their Fourier transforms, $\tilde H_1(x)$ and $\tilde H_2(x)$,  describing nonlocal component of spin density (and torque) in the absence of current.
$\tilde H_1$ has a finite adiabatic component ($q=0$, i.e., local component), while $\tilde H_2$ does not.
\label{FIGnlt1}
}
\end{figure}

In terms of $\stilpara$ and $\stilperp$, the equillibrium contribution (eq. (\ref{sequil})) is written as
\begin{eqnarray}
{\stil}^{\parallel {(0)}} (\xv,t) 
 &=&
\frac{1}{\Delta}\sumqv e^{-i\qv\cdot\xv}
\left[
\bar{A}_0^{\rm s} (\qv,\xv) H_1(\qv)+ \bar{A}_0^{\rm a} (\qv,\xv) H_2(\qv)
 \right]
\nonumber\\
{\stil}^{\perp {(0)}} (\xv,t) 
 &=&
\frac{1}{\Delta}\sumqv e^{-i\qv\cdot\xv}
\left[
\bar{A}_0^{\rm a} (\qv,\xv) H_1(\qv)- \bar{A}_0^{\rm s} (\qv,\xv) H_2(\qv)
 \right],
\end{eqnarray}
where the correlation functions are given by
\begin{eqnarray}
H_1(\qv) &=& \frac{\hbar\Delta}{V} \sum_{\kv\pm} 
 {\rm P}
\frac{f_{\kv\pm}}{\pm2\Delta+\frac{2\kv\cdot\qv+q^2}{2m}} 
\nonumber\\
H_2(\qv) &=& \frac{\hbar\Delta}{V} \sum_{\kv\pm} 
\frac{\pi}{2}(f_{\kv+}-f_{\kv-})
\delta\left(\pm2\Delta+\frac{2\kv\cdot\qv+q^2}{2m}\right),
\end{eqnarray}
and
$\bar{A}_\mu^{s}(\qv,\xv) 
  \equiv \half\sum_\pm e^{\mp i\phi(\xv)}A_\mu^\pm$ 
and
$\bar{A}_\mu^{a}(\qv,\xv) 
  \equiv \half\sum_\pm (\mp i)e^{\mp i\phi(\xv)}A_\mu^\pm$
(they depend on both $\qv$ and $\xv$).
These can be written as
$\bar{A}_\mu^{s}(\qv,\xv)= \evsph(\xv)\cdot \Av_\mu(\qv)$
and
$\bar{A}_\mu^{a} (\qv,\xv)= \evph(\xv)\cdot \Av_\mu(\qv)$, where
$\evsph\equiv (\cos\phi,\sin\phi,0)$.
Integration over $\kv$ is carried out to obtain
\begin{eqnarray}
H_1(\qv) &=& \se \tilde{H}_1(\qv)
\nonumber\\
\tilde{H}_1(\qv) & \equiv &
 \frac{3}{4}\frac{1}{(3+\zeta^2)}\frac{1}{\qtil^2}
\sum_{\pm} \left[
-\frac{1}{2\qtil}(\qtil^2-1)(\qtil^2-\zeta^2)
\ln\left|\frac{(\qtil+1)(\qtil\pm\zeta)}{(\qtil-1)(\qtil\mp\zeta)}\right|
\pm (1\pm\zeta)(\zeta\pm\qtil^2) \right],\nonumber\\
\end{eqnarray}
where $\qtil=|\qv|/(2\kf)$, 
$\se=\frac{\kf^3}{6\pi^2}\zeta(3+\zeta^2)$ is the electron spin density,
$\kf\equiv \half(\kf_++\kf_-)$, 
$\zeta\equiv \frac{\kf_+-\kf_-}{\kf_++\kf_-}$, 
and $\tilde{H}_1$ is normalized to be
$\tilde{H}_1=1+O(q^2)$.
We also obtain
\begin{eqnarray}
H_2(\qv) &=& 
\frac{\kf^3 \zeta}{16\pi}\frac{1}{\qtil}
\sum_{\pm}\int_{1-\zeta}^{1+\zeta} d\ktil \ktil 
\theta(\ktil\qtil -|\qtil^2\pm\zeta)|)
\equiv \kf^3 \tilde{H}_2(\qv),
\end{eqnarray}
where the integral is to be taken only in the regime
${\ktil>|\frac{1}{\qtil}(\qtil^2\pm\zeta)|}$.
Correlation functions, $\tilde H_1$ and $\tilde H_2$, are plotted in 
$q$-space and in real space in Fig. \ref{FIGnlt1}.

Similarly, current contributions are calculated as
\begin{eqnarray}
{\stil}^{\parallel {(1)}} (\xv,t) 
 &=&
\frac{1}{\Delta}\sumqv e^{-i\qv\cdot\xv}
\left[
(\Ev\cdot \bar{A}^{\rm s}) \chi_1(\qv)+ (\Ev\cdot \bar{A}^{\rm a}) \chi_2(\qv)
 \right]
\nonumber\\
{\stil}^{\perp {(1)}} (\xv,t) 
 &=&
\frac{1}{\Delta}\sumqv e^{-i\qv\cdot\xv}
\left[
(\Ev\cdot \bar{A}^{\rm a}) \chi_1(\qv) - (\Ev\cdot \bar{A}^{\rm s}) \chi_2(\qv)
 \right],  \label{stils}
\end{eqnarray}
where
\begin{eqnarray}
\chi_1(\qv) &=& \frac{e\tau\Delta}{6\pi m^2 V} \sum_{\kv\pm} 
\left( \kv\cdot\left(\kv+\frac{\qv}{2}\right)\right) 
i\frac{\gr_{\kv\pm}-\ga_{\kv\pm}}
 {\epsilon_{\kv+\qv}-\epsilon_{\kv}\pm2\Delta} 
\nonumber\\
\chi_2(\qv) &=& \frac{e\tau\Delta}{6\pi m^2 V} \sum_{\kv\pm} 
\left( \pm\frac{\pi}{2} \right) 
\left( \qv\cdot\left(\kv+\frac{\qv}{2}\right)\right) 
\delta( {\epsilon_{\kv+\qv}-\epsilon_{\kv}\pm2\Delta} )
i(\gr_{\kv\pm}-\ga_{\kv\pm}).
\label{chi2}
\end{eqnarray}

\begin{figure}[tbh]
  \begin{center}
  \includegraphics[width=7cm]{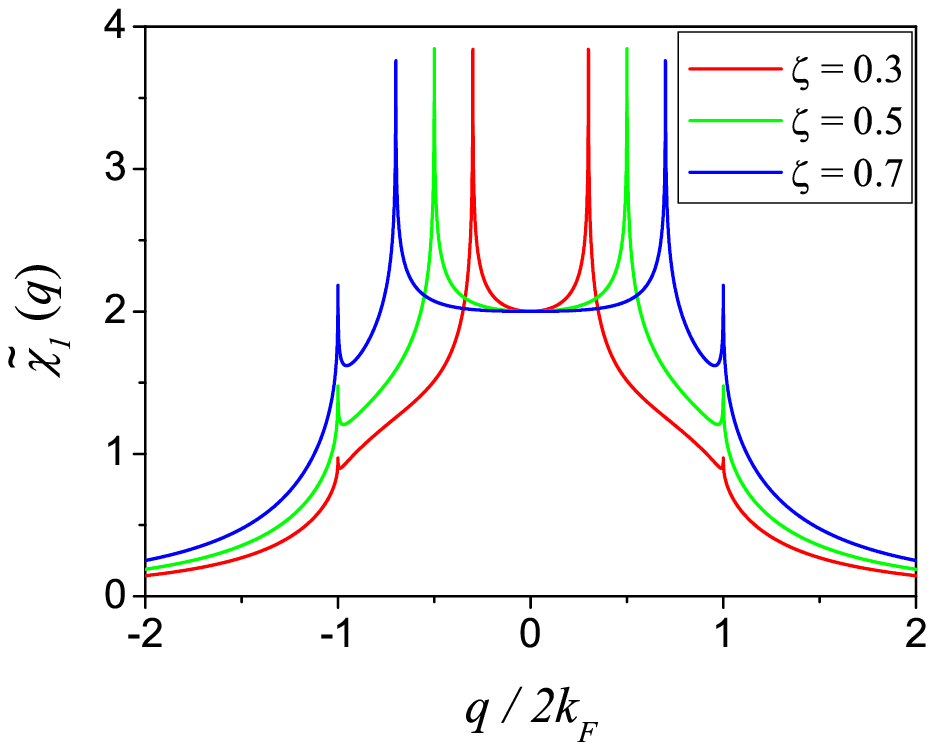}
  \includegraphics[width=7cm]{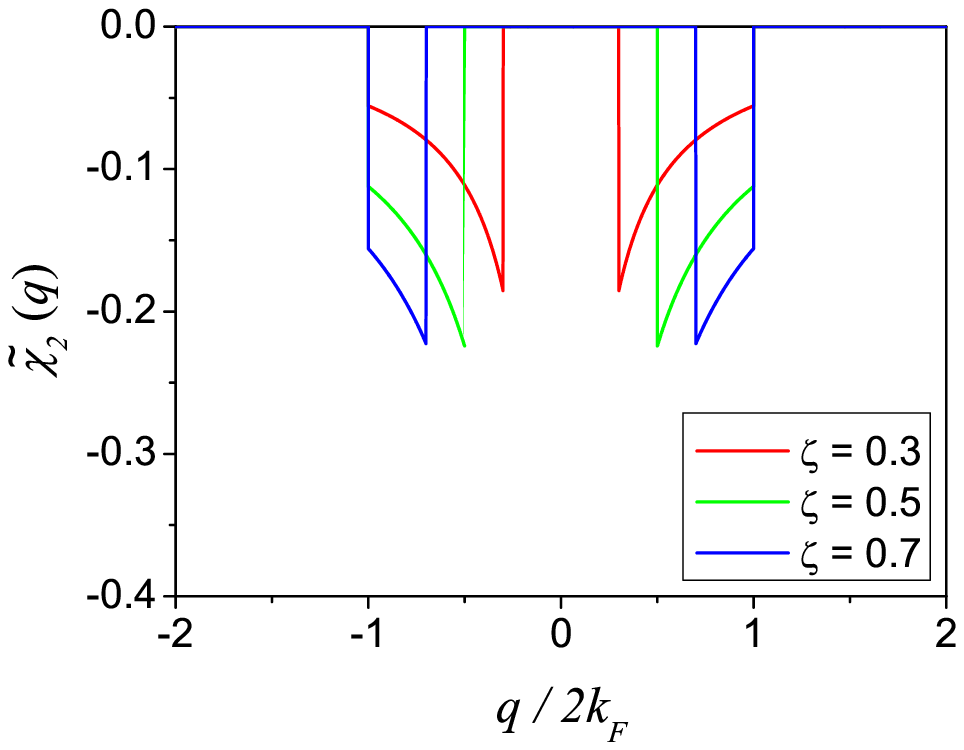}
  \includegraphics[width=7cm]{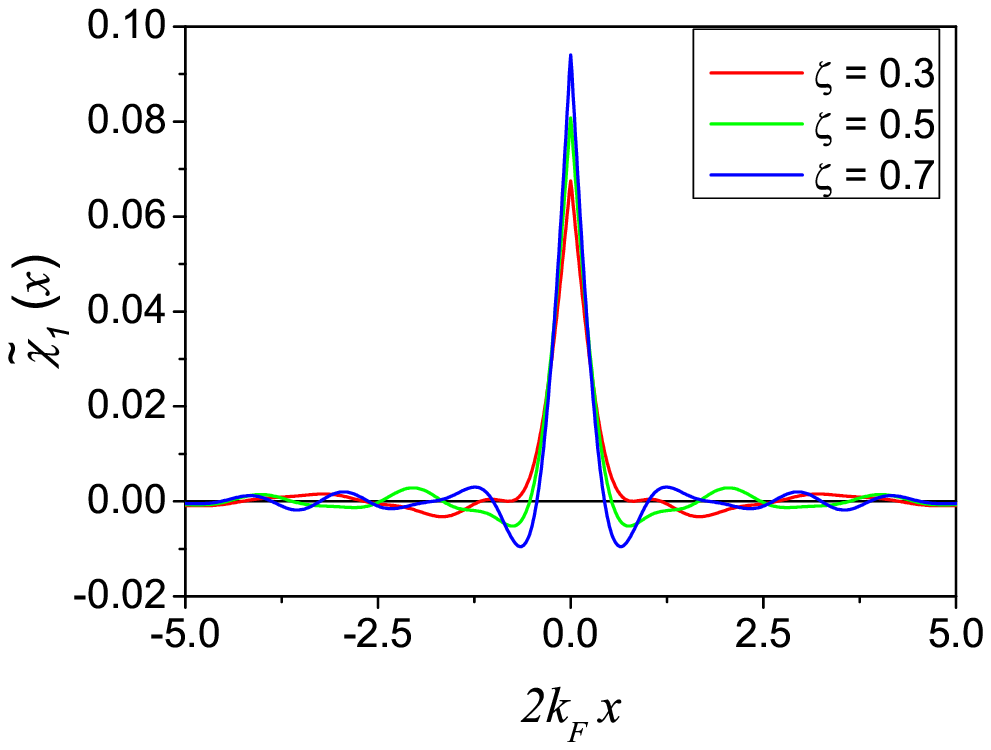}
  \includegraphics[width=7cm]{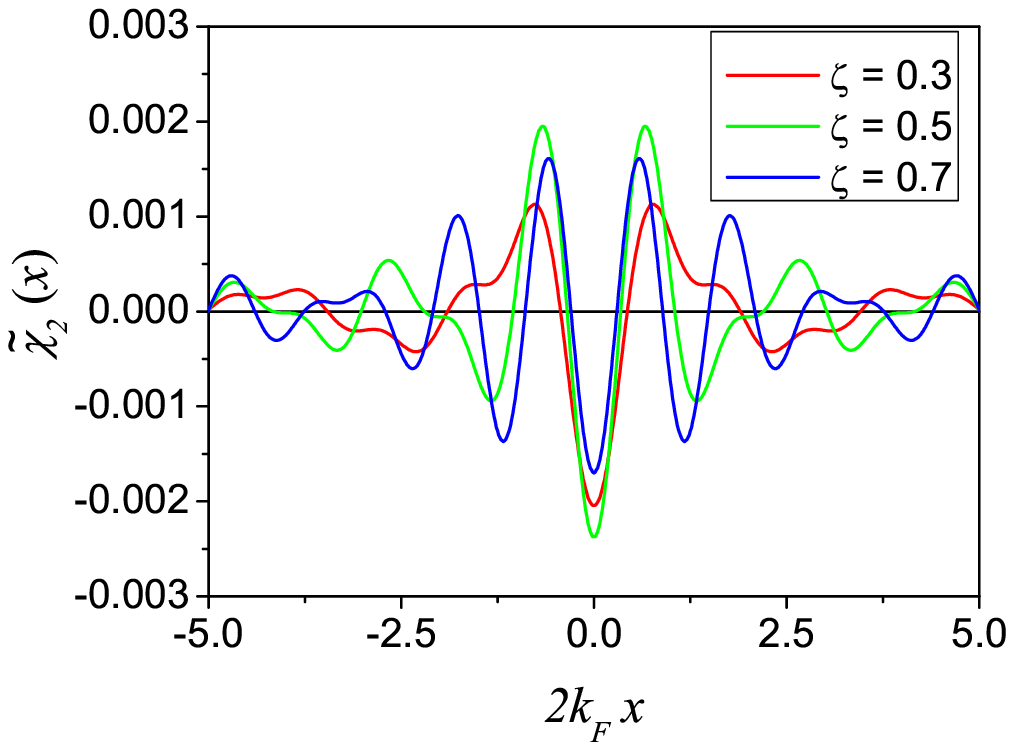}
  \end{center}
\caption{ (Color online)
Plot of correlation functions $\tilde\chi_1(q)$, $\tilde\chi_2(q)$ and
their Fourier transforms, $\tilde\chi_1(x)$ and $\tilde\chi_2(x)$,   describing nonlocal components of spin density and torque in the presence of current.
$\tilde\chi_1$ has a finite adiabatic component ($q=0$ i.e., local component) while $\tilde\chi_2$ does not.
\label{FIGnlt2}
}
\end{figure}

The summation over $\kv$ in eq. (\ref{chi2}) is carried out using
$\Vinv\sum_{\kv}=\int d\epsilon \DOS(\epsilon) \int_{-1}^{1}\frac{d\cos\theta_k}{2}$,
where $\theta_k$ is the angle of $\kv$ measured from the $\qv$ direction.
We carry out energy integration first, assuming that the dominant pole arises from the green function and the residue contribution from $\frac{1}{\epsilon_{\kv+q}-\epsilon_{\kv} \pm2\Delta} $ is neglected.
After some calculation, we otain
\begin{eqnarray}
\chi_1(\qv) &=& \frac{e\tau}{2m}(n_+-n_-) \tilde{\chi}_1(\qv)
\nonumber\\
\tilde{\chi}_1 &\equiv& \frac{1}{3+\zeta^2} \sum_{\pm}
(1\pm\zeta)\left[1+\frac{1+\zeta^2\pm\zeta-\qtil^2}{2(1\pm\zeta)\qtil}
\ln \left| \frac{(1+\qtil)(\qtil\pm\zeta)}{(1-\qtil)(\qtil\mp\zeta)} \right|
\right],
\end{eqnarray}
where $n_\pm=\frac{\kf_{\pm}^3}{6\pi^2}$ is the spin-resolved electron density.
$\tilde{\chi}$ is normalized to be $\tilde{\chi}=1+O(\qtil^2)$, and 
thus $\Ev \chi_1(\qv) =\frac{1}{e}\jsv \tilde{\chi}_1(\qtil)$
($\jsv\equiv \Ev(n_+-n_-)e^2\tau/m$ is spin current vector).
We similarly obtain
\begin{equation}
\chi_2(q) = -\frac{m^2\Delta^2}{12\pi n\kf}
\frac{\sigma_0}{e}\frac{\thetast(q)}{\qtil} 
\equiv \frac{\sigma_0}{e} \tilde{\chi}_2(q),
\end{equation}
where
\begin{equation}
\tilde{\chi}_2(q) = -\frac{\pi}{4}
\frac{\zeta^2}{1+3\zeta^2}\frac{\thetast(q)}{\qtil} ,
\end{equation}
and
\begin{equation}
\thetast(q) \equiv 
\left\{ 
 \begin{array}{cc}
1 & \kf_+-\kf_- \leq |q| \leq \kf_++\kf_- \\
0 & {\rm otherwise} \\
\end{array} \right.
\end{equation}
represents the regime of Stoner excitation, 
$\sigma_0=e^2n\tau/m$ is the Boltzmann conductivity, and $n=n_++n_-$ is the total electron density.
As is obvious, for small $\qtil$, $\chi_2=0$.
In the real space representation,
\begin{eqnarray}
\stil^{\parallel(1)} (\xv) 
&=&
\frac{1}{e\Delta}  \sum_{i}\intx'  \left[
{\js^i}
\tilde{\chi}_{1}(\xv-\xv') (\evsph(\xv)\cdot\Av_i(\xv'))
+{j^i}
\tilde{\chi}_2(\xv-\xv') (\evph(\xv)\cdot\Av_i(\xv') )
\right]\nonumber\\
\end{eqnarray} 
and
\begin{eqnarray}
\stil^{\perp(1)}(\xv) 
&=& 
\frac{1}{e\Delta}  \sum_{i}\intx'
\left[
{\js^i}  
\tilde{\chi}_{1}(\xv-\xv') (\evph(\xv)\cdot\Av_i(\xv'))
-{j^i}\tilde{\chi}_2(\xv-\xv') (\evsph(\xv)\cdot\Av_i(\xv') )
\right].\nonumber\\
\end{eqnarray}
Correlation functions, $\tilde\chi_1$ and $\tilde\chi_2$, are ploted in 
$q$-space and real space in Fig. \ref{FIGnlt2}.
Note that $\tilde{\chi}_1$- and $\tilde{\chi}_2$-terms are proportional to spin current and charge current, respectively, only in the adiabatic limit ($\qtil=0$), but are not necessarily so when nonadiabaticity sets in, since $\tilde{\chi}_1(\qtil)$ and $\tilde{\chi}_2(\qtil)$ can depend on polarization $\zeta$ in a complicated manner.

\section{Torque in Landau-Lifshitz-Gilbert Equation}

The electron spin density induces a torque on local spin at $\xv$, given by
\begin{equation}
\torquev(\xv)= -\frac{\Delta}{S}(\Sv\times \sev) 
= \Delta (-\stilperp \evth+ \stilpara \evph)
\label{torquefull}
\end{equation}
Namely, $\theta$- and $\phi$-components of torque are respectively given by
\begin{eqnarray}
\torque_\theta &=& -\Delta \ \stilperp \nonumber\\
\torque_\phi &=& \Delta \ \stilpara.
\end{eqnarray}
Thus, if one solves the Landau-Lifshitz(-Gilbert) equation of local spin taking account of eq. (\ref{torquefull}), one can include all the effects arising from conduction electrons.
From the above calculation, we now have full spin density as
$\stil^{\nu}=\stil^{\nu(0)} + \stil^{\nu(1)}$ ($\nu=\parallel,\perp$).
The torque acting on a spin at $\xv$ is then obtained as 
\begin{equation}
\torquev=\torquev^{(0)}+\torquev^{(1)},
\end{equation}
 where
$\torquev^{(0)}$ is an equillibrium part, 
\begin{eqnarray}
\torquev^{(0)}(\xv) &=& 
\intx' \left[ 
H_1(\xv-\xv') 
(\evph(\xv)(\evsph(\xv)\cdot \Av_0(\xv'))
-\evth(\xv)(\evph(\xv)\cdot \Av_0(\xv')))
 \right. \nonumber\\ && \left.
+H_2(\xv-\xv') 
(\evph(\xv)(\evph(\xv)\cdot \Av_0(\xv'))
+\evth(\xv)(\evsph(\xv)\cdot \Av_0(\xv')))
\right].
\end{eqnarray}
and the second part is due to current:
\begin{eqnarray}
\torquev^{(1)}(\xv) 
&=& 
 \sum_i \frac{\js^i}{e} \intx'
\left[ 
\tilde\chi_{1}(\xv-\xv') 
[\evth(\xv)(\evph(\xv)\cdot\Av_i(\xv'))
-\evph(\xv)(\evsph(\xv)\cdot\Av_i(\xv'))]
\right]
\nonumber\\
&& + 
 \sum_{i}  \frac{j_i}{e} \intx'
\left[ 
\tilde\chi_{2}(\xv-\xv') 
[\evth(\xv)(\evsph(\xv)\cdot\Av_i(\xv')) 
+\evph(\xv)(\evph(\xv)\cdot\Av_i(\xv') ]
\right]. \nonumber\\
&&
\end{eqnarray}

In the adiabatic limit, 
$H_1(\xv-\xv')\rightarrow \se\delta(\xv-\xv')$ and 
$\tilde{\chi}_1(\xv-\xv')\rightarrow \delta(\xv-\xv')$ become local, and 
$H_2$ and $\tilde\chi_2$ vanish. Thus, total torque 
reduces to a local one,
\begin{eqnarray}
\torquev_{\rm ad} (\xv)
&= & 
\left[\evth(\xv)\times \left(\evph(\xv)\times
\left[\se A_0(\xv)
+\sum_i\frac{\jsv}{e}\Av_i(\xv)\right]\right)\right]
\nonumber\\
&=& \half \left(\se\partial_t+\frac{a^3}{e}\jsv\cdot\nabla \right)\evs,
\label{torquead}
\end{eqnarray}
where we use
$(\evsph(\xv)\cdot\Av_\mu(\xv))=(\evth(\xv)\cdot\Av_\mu(\xv))
=A_\mu^\parallel$ and 
$(\evph(\xv)\cdot\Av_\mu(\xv))=A_\mu^\perp$.
When nonadiabaticity sets in, we see that 
(1). the torque becomes nonlocal, 
(2). torques in $\theta$- and $\phi$-directions become mixed, as seen by comparing $\chi_1$ and $\chi_2$.
The second role of mixing torque components is very important, since it means that the nonadiabaticity mixes spin-transfer torque ($\torque_z$) and force ($F$).
The oscillating torque found here is consistent with previous observations\cite{Waintal04,Xiao06}.
The oscillation is of the order of $2\kf$, as seen in Fig. \ref{FIGnlt2}.
Thus, this nonlocal torque is of quantum origin.
The classical estimate of conduction electron spin or torque (such as on the basis of the  Landau-Lifshitz(-Gilbert) equation) thus fails when nonadiabaticity is to taken into account.

Let us denote the torque above as
\begin{equation}
\torquev=\torquev_{\rm ad}+\torquev_{\rm nl},
\end{equation}
where $\torquev_{\rm nl}(\equiv \torquev-\torquev_{\rm ad})$ is purely nonlocal.
We here take account of torque arising from spin relaxation as considered in ref. \cite{KTS06},
\begin{equation}
\torquev_{\rm sf} = 
\frac{\alpha_{\rm sf}}{S} \Sv\times \dot{\Sv}
+\frac{a^3}{2eS}\beta_{\rm sf} 
(\Sv\times (\jsv\cdot\nabla)\Sv),
\end{equation}
where ${\alpha_{\rm sf}}$ and $\beta_{\rm sf}$ are proportional to the spin relaxation rate (see eqs. (\ref{eq:alpha1}) and (\ref{eq:beta1}) below).
(The spin relaxation process also produces nonlocal torque, but this contribution has so far not been addressed in metals, since it would be a nominal small correction. See also the discussion at the end of this section.)
The total torque due to electrons so far known is therefore summarized as
\begin{equation}
\torquev_{\rm tot}=\torquev_{\rm ad}+\torquev_{\rm nl}+\torquev_{\rm sf} ,
\end{equation}
and the effective equation of motion of local spin is given by
\begin{equation}
\partial_t \Sv= \frac{\alpha_0}{S} \Sv\times \dot{\Sv}+
\frac{g\muB}{\hbar}\Bv_{\rm eff}\times \Sv+\torquev_{\rm tot}, \label{Seq}
\end{equation}
where $\alpha_0$ and $\Bv_{\rm eff}$ are Gilbert damping and the effective field, both from nonelectron sources.
Using eq. (\ref{torquead}), we can rewrite eq. (\ref{Seq}) as
\begin{equation}
\left[\left(S+\frac{\se}{2}\right) \partial_t - \frac{a^3}{2e}\jsv\cdot\nabla \right] 
\evs= \frac{\alpha}{S} \Sv\times \dot{\Sv}
+\frac{a^3}{2eS}\beta_{\rm sf} (\Sv\times (\jsv\cdot\nabla)\Sv)
+\torquev_{\rm nl}+\frac{g\muB}{\hbar}\Bv_{\rm eff}\times \Sv,
\end{equation}
 where $\alpha\equiv \alpha_0+\alpha_{\rm sf}$.
We see here in the first term that the magnetization is now 
$S+\frac{\se}{2}$, i.e., the total spin, comprising local spin and conduction electron spin. 
The importance of using total spin instead of local spin in current-driven domain wall motion was stressed recently\cite{Barnes05}, but extra magnetization from conduction electrons is naturally taken account of in the present formalism (by a time component of the gauge field). 
We stress here that this effective Landau-Lifshitz-Gilbert equation 
(with $\beta$ and nonlocal torques) was mathematically derived. 
Thus Gilbert damping, which was recently argued to be incorrect in the presence of current\cite{Barnes06}, is justified.
As we see here, $\alpha$ and $\beta$ are finite and does not necessarily equal\cite{KTS06}, in contrast to the observation in ref. \cite{Tserkovnyak06}.
Very recent study\cite{Duine07} also supports non equal $\alpha$ and $\beta$..

Nonlocal oscillating torque around the domain wall in magnetic semiconductors was numerically studied quite recently\cite{Nguyen06}.
It was shown that the torque is asymmetric around the domain wall because of strong spin-orbit interaction
and that this interesting feature results in fast domain wall velocity.
Such an effect would be described by the spin-orbit correction to 
$\torquev_{\rm nl}$, but this has not been addressed so far in cases of metals (where spin-orbit interaction is caused by random impurities).

\section{Force}
Let us derive the expression for the force  acting on the spin structure.
The force is defined as the energy change when the structure is shifted.
Considering a displacement $\xv\rightarrow \xv+\Xv$,
the force due to electrons is\cite{Berger78,TK04}
\begin{eqnarray}
F_j & \equiv & -\deld{\Hex}{X_j}
  = -\frac{\Delta}{S} \intx \nabla_{j}\Sv \cdot\sev .
\end{eqnarray}
Using 
$\frac{1}{S}\nabla \Sv=\evth\nabla\theta+\evph \sin\theta\nabla\phi$
and $\evs\times\evth=-\evph$, $\evs\times\evph=\evth$,
we obtain
\begin{equation}
\Fv 
 =2\Delta \intx \left( \stilpara \Av^\perp -\stilperp \Av^\parallel \right).
\end{equation}
(This expression is equivalent to that described in ref. \cite{KTSS06}.)
From eqs. (\ref{stildef}) and (\ref{S1}) (noting 
$\gr_{\kv\sigma} \simeq -i\pi\delta(\epsilon_{\kv\sigma})$
on the $\kv$-integral) we obtain 
\begin{eqnarray}
\Fv 
&=&
-\sum_{\pm}\sum_{\kv\qv} \sum_{ij}  \frac{eE_i\tau\Delta}{m^2V}
 (\pm)\delta(\epsilon_{\kv\pm})
\nonumber\\ && \times 
\left[ 
 k_i \left(k+\frac{q}{2}\right)_j \frac{-i}
{\epsilon_{\kv+q}-\epsilon_{\kv} \pm2\Delta} 
\left( 
 A_j^\pm(\qv)  \Av^\mp(-\qv)
 - A_j^\mp(\qv) \Av^\pm(-\qv)
\right)
\right. \nonumber\\
&& \left.
-\pi q_i \left(\kv+\frac{q}{2}\right)_j 
\delta({\epsilon_{\kv+q}-\epsilon_{\kv} \pm2\Delta})
\left( 
 A_j^\pm(\qv) \Av^\mp(-\qv) + A_j^\mp(\qv)\Av^\pm(-\qv)
\right)
\right].
\label{F}
\end{eqnarray}
The second term vanishes in the adiabatic limit ($q=0$) while the first term remains finite.
We now demonstrate that the second term represents the reflection of the electron due to the spin texture.
In fact,  we find, by using
$\epsilon_{\kv+\qv}-\epsilon_{\kv}=\frac{1}{m}\qv\cdot(\kv+\frac{\qv}{2})$,
that the second term (we call $F^{\rm (na)}$) is 
\begin{eqnarray}
F^{\rm (na)}_j
&=&
-\sum_{\pm}\sum_{\kv\qv} \sum_{i}  \frac{4\pi eE_i\tau\Delta^2}{mV}
\delta(\epsilon_{\kv\pm}) 
\delta({\epsilon_{\kv+q}-\epsilon_{\kv} \pm2\Delta})
 A_i^\pm(\qv)
A_j^\mp(-\qv). \label{Fna}
\end{eqnarray}
We can show that this force is proportional to the resitivity due to spin structure, which is obained, on the basis of Mori formula, as (in the case of current along the $\xw$-direction)
\begin{equation}
\rhos= \frac{4\pi \Delta^{2}}{e^{2}n^{2}}\frac{1}{V}
 \sum_{\kv q \sigma} |A_\xw^\sigma(\qv)|^{2} 
\delta(\epsilon_{\kv+q,-\sigma} -\epsilon_{\kv,\sigma} ) \delta(\epsilon_{\kv,\sigma} ).
\label{rhos}
\end{equation}
In fact, 
\begin{equation}
\Fv^{\rm (na)}= \frac{e^3 \Ev \tau}{m}\rhos n^2 V 
  =e\Ne\rhos \jv, \label{Fandj}
\end{equation}
where $\Ne=nV$ is the total electron number.
This is the result in ref. \cite{TK04} extended to a general spin structure.
The derivation of resistivity using the Mori formula is shown in \S \ref{SECmori}.

\begin{figure}[tbh]
  \begin{center}
  \includegraphics[scale=0.3]{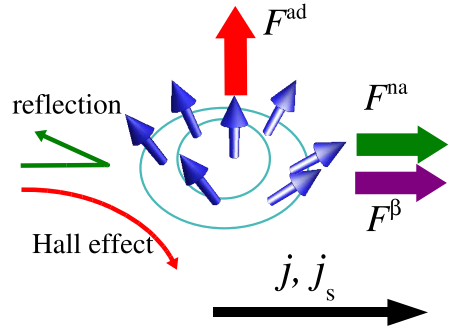}
\caption{ (Color online)
Schematic summary of three forces acting on spin structures. 
Electron reflection pushes the structure along the current flow ($F^{\rm na}$), and Hall effect due to spin chirality results in a force in the perpendicular direction ($F^{\rm ad}$). 
Spin relaxation results in a force along the direction of current ($F^{\beta}$).
\label{FIGvortexwall}}
  \end{center}
\end{figure}

If reflection is included in $F^{\rm (na)}$, what then is the origin of the first term of eq. (\ref{F})?
This becomes clear if we consider the adiabatic limit.
In fact, in this limit, the first term reduces to
\begin{eqnarray}
F_j^{\rm (ad)} 
&=&
\sum_{\pm}\sum_{\kv\qv} \sum_{i}  \frac{eE_i\tau}{2m}
n_{\pm} i \delta(\epsilon_{\kv\pm})
k^2 \left(
A_i^\pm(\qv) A_j^\mp(-\qv)
+ A_i^\mp(\qv) A_j^\pm(-\qv)
\right) 
\nonumber\\
&=&
i \frac{{\js}_i}{e} \intx \left(
A_i^+(\xv) A_j^-(\xv)
- A_i^-(\xv) A_j^+(\xv)
\right) .
\end{eqnarray}
By using
\begin{equation}
\left(
A_i^+(\xv) A_j^-(\xv)
- A_i^-(\xv) A_j^+(\xv)
\right) 
= -i \half \sin\theta 
(\partial_i\theta \partial_j\phi-\partial_j\theta \partial_i\phi),
\end{equation}
and
\begin{equation}
\partial_i\Sv \times \partial_j\Sv =\evs S^2 \sin\theta 
(\partial_i\theta \partial_j\phi
  -\partial_j\theta \partial_i\phi),
\end{equation}
we see that
\begin{eqnarray}
F_j^{\rm ad} 
&=& \frac{1}{2S^3} \sum_i \frac{{\js}_i}{e} \intx 
\Sv\cdot(\partial_i\Sv \times \partial_j\Sv )
=2\pi \sum_i \frac{{\js}_i}{e} \Phi_{ij},
\label{Fad}
\end{eqnarray}
where
\begin{equation}
\Phi_{ij}\equiv  \frac{1}{4\pi S^3}  \intx
\Sv\cdot(\partial_i\Sv \times \partial_j\Sv )
\end{equation}
is a vortex number defined in three dimensions.
In the case of a thin system, this reduces  to 
\begin{equation}
\Phi_{ij}=\nvortex \thickness,
\end{equation}
where $\thickness$ is the thickness of the system,
and
\begin{equation}
\nvortex \equiv  \frac{1}{4\pi S^3}  \int d^2x
\Sv\cdot(\partial_i\Sv \times \partial_j\Sv ),
\end{equation}
is a topological number. 
This force is, in fact, a back reaction of the Hall effect due to spin chirality\cite{Ye99,TKawamura02,OTN04}, and was derived first phenomenologically by Berger\cite{Berger86}, then rigorously in ref. \cite{KTSS06}, and also
assuming a vortex structure in ref. \cite{SNTKO06}.
Note that the adiabatic force is included in (adiabatic) spin transfer torque, while nonadiabatic force is not.
The two forces are schematically described in Fig. \ref{FIGvortexwall}.

\section{Resistivity and Hall Resistivity Calculated Using Mori Formula}
\label{SECmori}

The resistivity of the domain wall was calculated by Cabrera and Falicov\cite{Cabrera74} and later by others\cite{TF97,Levy97,TZMG99,GT01}.
Here, we derive the resistivity (both diagonal and off-diagonal (i.e., Hall) components) due to the general spin texture on the 
basis of the linear response theory (Mori formula\cite{Mori65,Gotze72})\cite{GT01} in the clean limit of $\tau\rightarrow\infty$.
The Mori formula relates the resistivity $\rhoS_{ij}$ ($ij$ being spatial directions)
to the correlation of random forces in the weak scattering case as
\begin{equation}
{\rhoS}_{ij}=\left(\frac{e^{2} n}{m}\right)^{-2}\lim_{\omega\rightarrow 0}
\frac{1}{\hbar\omega} 
{\rm Im}[\chi_{\dot{J}_i\dot{J}_j}(\hbar\omega)-\chi_{\dot{J}_i\dot{J}_j}(0)].
\label{Mori}
\end{equation}
Here, $\chi_{\dot{J}_i\dot{J}_j}(i\omega_{\ell})\equiv
-(\hbar/\beta V) <\dot{J}_i(i\omega_{\ell})\dot{J}_j(-i\omega_{\ell})>$
with $\dot{J_i}\equiv dJ_i/dt= \frac{ i}{\hbar}[H,J_i]$,
where $H$ is the total Hamiltonian and 
$J_i \equiv \frac{e\hbar}{m} \sum_\kv k_i \cdag_\kv c_\kv$ is the total current.
The correlation function 
$\chi_{\dot{J}\dot{J}}(\hbar\omega)$ in eq. (\ref{Mori}) denotes 
an analytical continuation of the correlation function calculated for 
imaginary frequency, 
i.e., $\chi_{\dot{J}\dot{J}}(\hbar\omega)\equiv 
\chi_{\dot{J}\dot{J}}(i\omega_{\ell}\rightarrow \hbar \omega+i0)$.
\begin{figure}[tbh]
  \begin{center}
  \includegraphics[scale=0.3]{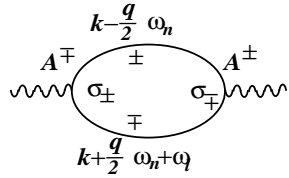}
\caption{Diagram representing the resistivity in Mori formula at the lowest order in the gauge field.
\label{FIGmori}}
  \end{center}
\end{figure}

The nonconservation of the current (i.e., finite $\dot{J}$) arises from the 
scattering by the spin texture. 
In fact, eq. (\ref{Le}) leads to 
\begin{eqnarray}
\dot{J_{i}} &=&  i\left(\frac{e}{m}\right)
\sum_{\kv,\qv} \left[ -2\Delta  \sum_{\sigma} A_{i}^{-\sigma}  \adag_{\kv+\qv}\sigma_{\sigma} a_{\kv} \right.\nonumber\\
 && \left. +
 \frac{\hbar^2}{2m} 
 \left( 
A_{i}^{\alpha}((2\kv+\qv)\cdot \qv)  - q_i (\Av^{\alpha}\cdot(2\kv+\qv))
-\hbar q_i A_0^{\alpha} \right) 
\adag_{\kv+\qv}\sigma_{\alpha} a_{\kv} \right],
\end{eqnarray}
where we neglect higher order terms in $A$.
We consider a static spin texture (i.e., $A_0=0$), 
and then
\begin{equation}
\dot{J_{i}} = -i2\left(\frac{e}{m}\right) \Delta
\sum_{\kv,\qv} \sum_{\sigma} A_{i}^{-\sigma}  
\adag_{\kv+\qv}\sigma_{\sigma} a_{\kv} .
\end{equation}
The Fourier transform in imaginary time $\tau$, defined as
$\dot{J_{i}}(i\omega_{\ell})\equiv \int_0^{\beta} e^{-i\omega_\ell \tau}
 \dot{J_{i}}(\tau)$, 
where $\omega_{\ell}\equiv \frac{2\pi\ell}{\beta}$ is 
a bosonic thermal frequency,
is given as
\begin{equation}
\dot{J_{i}}(i\omega_{\ell}) = -i2\left(\frac{e}{m}\right) \Delta
\sum_{\kv,\qv,n} \sum_{\sigma} A_{i}^{-\sigma}  
\adag_{\kv+\qv,n+\ell}\sigma_{\sigma} a_{\kv,n} ,
\end{equation}
where 
$a_{\kv,n}\equiv \frac{1}{\sqrt{\beta}} 
 \int_0^{\beta} e^{i\omega_n \tau} a_{\kv}(\tau)d\tau$ and $\omega_n$ 
represents the fermionic frequency $\omega_n\equiv \frac{\pi(2n-1)}{\beta}$.
The thermal correlation function is then obtained as
\begin{equation}
\chi_{\dot{J}_i\dot{J}_j}(i\omega_{\ell}) =
-{\hbar} \left(\frac{2e\Delta}{m}\right)^{2} \frac{1}{V}\sum_{kq\sigma} 
 A_i^{\sigma}(\qv) A_j^{-\sigma}(-\qv)
\frac{1}{\beta}\sum_{n}G_{k+q,n+\ell,-\sigma} G_{kn\sigma},
\label{chimori}
\end{equation}
where the imaginary time Green function is defined as
\begin{equation}
G_{kn\sigma} =\frac{1}{i\omega_n-\ekvs}.
\end{equation}
The summation over $\omega_{n}$ is carried out to obtain
\begin{equation}
\chi_{\dot{J}_i\dot{J}_j}(i\omega_{\ell})=
-\hbar\left(\frac{2e\Delta}{m}\right)^{2} \frac{1}{V}\sum_{kq\sigma}
 A_i^{\sigma}(\qv) A_j^{-\sigma}(-\qv)
\frac{
  f(\epsilon_{k+q,-\sigma})- f(\epsilon_{k,\sigma})  }
  { \epsilon_{k+q,-\sigma} -\epsilon_{k,\sigma} -i\omega_{\ell} }.
	\label{chidotcor}
\end{equation}
Thus the resistivity is obtained as
\begin{equation}
{\rhoS}_{ij}=
\lim_{\omega\rightarrow0} \frac{4\Delta^2}{e^2n^2} \frac{1}{V}\sum_{\kv\qv\sigma} \Im \left[
 A_i^{\sigma}(\qv) A_j^{-\sigma}(-\qv)
\frac{1}{\omega} \frac{
  f(\epsilon_{k+q,-\sigma})- f(\epsilon_{k,\sigma})  }
  { \epsilon_{k+q,-\sigma} -\epsilon_{k,\sigma} -\omega-i0 }\right].
	\label{rhosij}
\end{equation}
Then, choosing the current direction as $\xw$, resistivity is rewritten as 
\begin{equation}
\rhoS \equiv {\rhoS}_{\xw\xw}
= \frac{4\pi \Delta^{2}}{e^{2}n^{2}}\frac{1}{V}
 \sum_{\kv q \sigma} |A^\sigma_\xw(\qv)|^{2} 
\delta(\epsilon_{\kv+q,-\sigma} -\epsilon_{\kv,\sigma} ) \delta(\epsilon_{\kv,\sigma} ).
\label{rhocclean}
\end{equation}
Thus, the reflection force of eq. (\ref{Fna}) is proportional to the resistivity due to spin, as we explained for eq. (\ref{Fandj}).
The $\kv$-summation can be carried out easily as
\begin{equation}
\Vinv \sum_{\kv} 
\delta(\epsilon_{\kv+q,-\sigma} -\epsilon_{\kv,\sigma} ) \delta(\epsilon_{\kv,\sigma} )
=\frac{\DOS_\sigma m}{2 k_{F\sigma}q}\thetast(q),
\end{equation}
where $q\equiv |\qv|$.

Resistance, 
$\RS\equiv \frac{L}{A}\rhoS $,
due to spin texture is related to the reflection probability $R$, according to four-terminal Landauer-B\"uttiker formula\cite{Buttiker85}, as 
$\RS=\frac{\pi\hbar}{e^2} \frac{R}{1-R}$, and hence eq. (\ref{Fandj}) 
indeed relates the force to the reflection of the electron.
Equation (\ref{Fandj}) can also be written, using the density of states 
$\DOSV=\frac{m \kf V}{2\pi^2 \hbar}$ (neglecting spin splitting), as
\begin{equation}
\Fena= \frac{1}{3} eV_{\rm S} \DOSV {2\kf\hbar}\frac{\kf\hbar}{mL},
\end{equation}
where $V_{\rm S}\equiv \RS I$ is a voltage drop by spin texture and  $I\equiv Aj$ is current ($A$ being the cross section).
This equation clearly indicates that the force is due to the momentum transfer
(${2\kf\hbar}$ per electron, with frequency of $\frac{\kf\hbar}{mL}$)  
multiplied by the number of electrons that contribute to the resistance
($\frac{1}{3}eV_{\rm S} \DOSV$).

In contrast to the diagonal component of resistivity, which vanishes in the adiabatic limit, the off-diagonal (i.e., Hall) component remains finite in this limit.
In fact, we find 
\begin{eqnarray}
{\rhoS}_{ij}^{\rm ad} &=&
\lim_{\omega\rightarrow0} \left(\frac{4\Delta^2}{e^2n^2}\right) \frac{1}{V}\sum_{\kv\qv\sigma} \Im \left[
 A_i^{\sigma}(\qv) A_j^{-\sigma}(-\qv)
\frac{1}{\omega} \frac{
  f(\epsilon_{k,-\sigma})- f(\epsilon_{k,\sigma})  }
  { 2\sigma \Delta-\omega }\right]
\nonumber\\
&=&
-\frac{1}{\hbar e^2n^2V}\sumqv \Im \left[
 A_i^{\sigma}(\qv) A_j^{-\sigma}(-\qv) \right]
\sum_{\kv\sigma} f(\epsilon_{k,\sigma})
\nonumber\\
&=&
\frac{n_+-n_-}{2\hbar e^2n^2V }\intx 
\evs\cdot(\partial_i\Sv \times \partial_j\Sv )
=\frac{2\pi S^2 P}{\hbar e^2 n V}\Phi_{xy}. \label{rhoxy}
\end{eqnarray}
(Correctly, we have retained the antisymmetric component, $\half({\rhoS}_{ij}-{\rhoS}_{ji})$, in the second line.)
This is the Hall effect caused by the spin chirality, or spin Berry phase\cite{Ye99}, demonstrated in the slowly varying case\cite{TKawamura02,OTN04}.
Comparing eqs. (\ref{Fad}) and (\ref{rhoxy}), we see that the force in the adiabatic limit is exactly due to the Hall effect from spin chirality:
\begin{eqnarray}
F_j^{\rm ad} 
&=& \frac{2e^2n^2}{n_+-n_-} \sum_i {\js}_i {\rhoS}_{ij} 
\nonumber\\
&=& e^2n \sum_i j_i {\rhoS}_{ij} .
\end{eqnarray}


\section{Spin Relaxation}
Let us briefly look into the role of spin relaxation discussed in ref. \cite{KTS06}.
The spin flip scattering was introduced there as
\begin{equation}
H_{\rm sf} = u_s \intx \sum_i \Sv_i \delta(\xv-\Rv_i)
(\cdag\sigmav c)_\xv,
\end{equation}
where $\Sv_i$ represents an impurity spin at site $\Rv_i$.
(The spin-orbit interaction leads to essentially the same results as the spin flip case here.)
The quenched average for the impurity spin was taken to be 
\begin{eqnarray}
  \overline{S_i^\alpha S_j^\beta} 
&=& \delta_{ij} \delta_{\alpha\beta} \times \left\{ \begin{array}{cc} 
    \overline{S_\perp^2} & (\alpha, \beta = x,y) \\ 
    \overline{S_z^2}     & (\alpha, \beta = z) 
    \end{array}. \right.
\end{eqnarray}
Kohno {\it et al.}\cite{KTS06} considered the adiabatic limit, and obtained the torque due to spin flip scattering as
\begin{equation}
\torquev_{\rm sf} = 
\frac{\alpha_{\rm sf}}{S} \Sv\times \dot{\Sv}
+\frac{a^3}{2eS}\beta_{\rm sf} 
(\Sv\times (\jsv\cdot\nabla)\Sv),
\end{equation}
where coefficients are given by
\begin{eqnarray}
  \alpha_{\rm sf} 
&=&  \pi n_{\rm s} u_{\rm s}^2 \left[\, 
     2 \overline{S_z^2} \DOS_+ \DOS_- 
    + \overline{S_\perp^2} (\DOS_+ ^2 + \DOS_-^2 ) 
    \right] . 
\label{eq:alpha1}
\\
  \beta_{\rm sf} 
 &=&  \frac{\pi n_{\rm s}u_{\rm s}^2}{\Delta} 
    \left[ 
    \bigl( \overline{S_\perp^2} + \overline{S_z^2} \bigr) \DOS_+ 
  + \frac{1}{P_j} 
    \bigl( \overline{S_\perp^2} - \overline{S_z^2} \bigr) \DOS_- 
    \right] ,
\label{eq:beta1}
\end{eqnarray}
where $\DOS_\pm\equiv \frac{m\kf_{\pm}}{2\pi^2}$ 
and $n_s$ denotes the concentration of magnetic impurity.
The spin relaxation process also produces nonlocal torque, but this contribution has not been  addressed so far, since it is a nominal small correction in the present context.

The torque, $\torquev_{\rm sf}$, is due to the spin polarization caused by spin relaxation ($\torquev_{\rm sf}=-\frac{\Delta}{S}\Sv\times\sev_{\rm sf}$), 
\begin{equation}
\sev_{\rm sf} = -\frac{1}{\Delta a^3}
\left( \alpha_{\rm sf}\dot{\Sv} +\frac{a^3}{2eS}\beta_{\rm sf} 
(\jsv\cdot\nabla)\Sv \right).
\end{equation}
This spin density gives rise to a force,
\begin{equation}
F^{\rm sf}_i = \frac{1}{S}\sumx \nabla_i \Sv \cdot
\left( \alpha_{\rm sf}\dot{\Sv} +\frac{a^3}{2eS}\beta_{\rm sf} 
(\jsv\cdot\nabla)\Sv \right).
\end{equation}
The force due to the $\beta$ term is thus given as\cite{Thiaville05,Zhang04,KTSS06}
\begin{equation}
\Fv^{\beta}=\frac{1}{2eS} \gamma\beta_{\rm sf} \jsv,
\end{equation}
where $\gamma\equiv \intx (\nabla_i \Sv)^2$, 
with $\hat{\iv}\parallel\jsv$.
If we consider a quasi-two-dimensional system, the total force due to current, $\Fv=\Fv^{\rm (na)}+\Fv^{\rm (ad)}+\Fv^{\beta}$, is summarized as
\begin{equation}
\Fv=\jv
\left( e\Ne \rhoS + P\gamma \frac{1}{2eS}\beta_{\rm sf} \right)
+P \frac{a^3}{2eS} \jv\times \gv, \label{totalF}
\end{equation}
where $P=\js/j$ is the polarization of current and
$\gv\equiv \frac{4\pi S}{a^3}\Phi_{xy}\evz$
is the gyrovector of vorticity\cite{He06,SNTKO06}.
We see that the two forms of nonadiabaticity, due to fast spin texture and spin relaxation, both result in a force in the same direction, that of current.


\section{Motion of a Vortex}
Let us consider specifically the case of a single vortex 
in a film in a $xy$-plane.
As was explained by Shibata {\it et al.}\cite{SNTKO06}, the Lagrangian of a single vortex in terms of its position, $\Xv=(x,y)$, is given as
\begin{equation}
L_{\rm v}=\frac{1}{2}\gv\cdot(\dot{\Xv}\times\Xv)-U(\Xv),
\end{equation}
where $g=\frac{4\pi \hbar S}{a^3}\Phi_{xy}$, and
$U$ denotes all the potential energy, including the effect of current. 
(Note that they considered the effect of current in the adiabatic limit separately from $U$ as a spin-torque term.)
The equation of motion is thus given by\cite{SNTKO06}
\begin{equation}
\dot{\Xv}=\frac{1}{g}\evz\times(\Fv-\tilde\alpha \dot{\Xv}),
\label{veq1}
\end{equation}
where $\Fv\equiv \frac{\partial}{\partial \Xv}U(\Xv)$ and
$\tilde\alpha\equiv \alpha\frac{\hbar S}{a^3} \thickness
\int d^2 x [(\nabla\theta)^2+\sin^2\theta (\nabla\phi)^2]$.
As seen, vortex has a velocity perpendicular to the applied force, $\dot{\Xv}\perp\Fv$, which might sound strange but is an interesting feature of a topological object.
The force due to current is given by eq. (\ref{totalF}), and so
eq. (\ref{veq1}) reduces to
\begin{equation}
\dot{\Xv}=\frac{a^3}{2eS}\jsv
+ \evz\times
\left(\frac{f}{e}\jv
-\frac{a^3\tilde\alpha}{4\pi S \Phi_{xy}}
 \dot{\Xv} \right),
\label{veq2}
\end{equation}
where 
$f=\frac{a^3}{8\pi S^2\Phi_{xy}}
( 2Se^2 \Ne\rhos+\gamma P \beta_{\rm sf})$.
The perpendicular force on the vortex of the adiabatic origin ($\jv\times \gv$ in eq. (\ref{totalF}))
is thus simply a spin torque, which induces the vortex to flow in the current direction.
This is easily understood since spin torque and the time-derivative term of the Landau-Lifshitz equation are combined into a Lagrange derivative along current flow, 
$\partial_t \rightarrow 
\left( \partial_t-\frac{1}{2eS}\jsv\cdot\nabla \right)$, 
which indicates that spin transfer torque drives any spin structure along the spin current.

Let us consider a single vortex in a film in more detail.
The Hamiltonian of local spin is modeled as
\begin{equation}
H_v=\frac{S^2\thickness}{2}\int  d^2x [
J[(\nabla\theta)^2+\sin^2\theta(\nabla\phi)^2]
+\Kp\cos^2\theta].
\end{equation}
Since we cannot obtain the vortex solution analytically, we approximate it as
\begin{equation}
\phi=\tan^{-1}\frac{y}{x}\pm\frac{\pi}{2}, \;\;\; 
\theta=
\left\{ \begin{array}{cc}
\frac{\pi}{2} & r> \lamv \\
\frac{\pi }{2(1-e^{-1})} (1-e^{-r^2/\lamv^2})  & r \leq \lamv
\end{array} \right.,
\end{equation}
where $r\equiv\sqrt{x^2+y^2}$ and
$\lamv$ is the size of the vortex core.
Then the gauge field is given as
$A_i^\pm\sim \pm i \frac{x_i(x\pm i y)}{\lamv^2 r} e^{-r^2/\lamv^2} \theta(\lamv-r)$, neglecting the small contribution from outside the core.
Choosing the current direction as $x$, we obtain
the Fourier transform as ($A$ is the area of the film)
\begin{equation}
A_x^\pm(\qv)=\pm i \pi\frac{\lamv}{2A} (\sqrt{\pi}(1-2q^2\lamv^2)+2iq\lamv)
 e^{-\frac{\lamv^2 q^2}{4}}.
\end{equation}
The resistivity due to the core is then calculated as
\begin{eqnarray}
\rho_{\rm v} &\simeq&  \frac{\pi^2}{2}
\frac{m^2 \kf \Delta^2 \lamv^2}{e^2 n^2 A}
 e^{-2\zeta^2 \kf^2\lamv^2 },
\end{eqnarray}
where we used the fact that resistivity is dominated by the contribution from $q\lesssim k_{F+}-k_{F-}=2\kf\zeta$.
The nonadiabatic force is then given as
\begin{equation}
F^{\rm na} \simeq \frac{j}{e}\hbar \left(\frac{\Delta}{\eF}\right)^2 
(\kf\lamv)^2 d  e^{-2\zeta^2 \kf^2\lamv^2 }.
\end{equation}
On the other hand, the force due to the topological Hall effect is given by
\begin{equation}
F^{\rm ad} \simeq \frac{j}{e}\hbar 
\end{equation}
for a vortex with $\nvortex=1$.


Let us estimate the magnitude of the Hall effect.
Hall resistivity is given as
\begin{equation}
\rhoxy\simeq  \frac{2\pi S^2 P \nvortex}{\hbar e^2 n A}.
\end{equation}
The Hall conductivity is given by
$\sigma_{xy}=\sigma_0
 \frac{\rho_{xy}/\rho_0}{1+(\rho_{xy}/\rho_0)^2}$ 
( $\sigma_0=\rho_0^{-1}$ is the Boltzmann conductivity) and hence
the ratio of the Hall current to applied current 
is obtained as 
\begin{equation}
\frac{j_\perp}{j_0} =  {\rho_{xy}/\rho_0}.
\end{equation}
Thus, the deviation of the electric current becomes significant if 
the ratio $ \frac{\rho_{xy}}{\rho_0} 
\sim 2\pi S^2 P\nvortex \frac{\ell}{\kf A}$ is of the order of unity.
The deviation of current because of the Hall effect would be important in clean samples.


In conclusion,
we studied the nonadiabatic correction to the torque acting on local spin arising from the electric current, treating conduction electrons fully quantum mechanically.
The nonadiabaticity we considered here is that due to the fast-varying spin texture. 
(The spin relaxation of conduction electrons results in the deviation of electron spin from  adiabaticity, and the effect is sometimes also called nonadiabaticity. This effect was considered in our preceeding paper\cite{KTS06}.)
The nonadiabatic torque was shown to be nonlocal in space, to have an oscillation with a period of the order of $2\kf$, as was argued by Waintal and Viret\cite{Waintal04} and Xia {\it et al.}\cite{Xiao06}, and to be of quantum origin.
This oscillation torque is represented as a force acting on the whole spin texture, and thus it represents the effect of momentum transfer\cite{Berger78,TK04}. 
Note that this nonlocal torque is totally different from the local $\beta$-term torque, which was  originaly introduced to simulate momentum transfer force\cite{Thiaville05}, although they both contribute to a force.
We found another type of force, topological or adiabatic force,  which remains finite in the adiabatic limit and is proportional to the vortex number.
These two forces were shown to be the counteraction of electron transport properties such as resistivity and the Hall effect. 

The authors are grateful to Y. Yamaguchi, T. Ono, M. Yamanouchi, H. Ohno, M. Kl\"aui, Y. Nakatani, A. Thiaville, A. Brataas, R. Egger, M. Thorwart, J. Ieda, J. Inoue, S. Maekawa and H. Fukuyama for valuable discussion.
G.T. is grateful to the Center of Advanced Study, Oslo, for its hospitality during his stay.

\appendix

\section{Details of calculation of spin density }

\label{SECspindensity}
We present, in this section, details of the calculation of spin density.
Equation (\ref{sSE}) is evaluated to the lowest order in $\Omz$ as
\begin{eqnarray}
\stil^{\pm}(\xv,t)^{\rm (A)} & = &
S_1+S_2+S_3+S_4+S_5
\nonumber\\
S_1 &=& \sum_{\omega\kv\qv} \sum_i B_i f'(\omega) 
\nonumber\\
&& \times 
\left[ \left(k+\frac{q}{2}\right)_i 
 \gr_{\kv-\frac{q}{2},\mp,\omega} 
 \gr_{\kv+\frac{q}{2},\pm,\omega} 
 \ga_{\kv+\frac{q}{2},\pm,\omega}
+\left( k-\frac{q}{2}\right)_i 
\gr_{\kv-\frac{q}{2},\mp,\omega}
\ga_{\kv-\frac{q}{2},\mp,\omega} \ga_{\kv+\frac{q}{2},\pm,\omega}
\right]
\nonumber\\
S_2
&=& -\sum_{\omega\kv\qv} \frac{B_0}{\Omz} f(\omega) 
\left(
\ga_{\kv-\frac{q}{2},\mp,\omega}
\ga_{\kv+\frac{q}{2},\pm,\omega}
- \gr_{\kv-\frac{q}{2},\mp,\omega}
\gr_{\kv+\frac{q}{2},\pm,\omega}
\right)
\nonumber\\
S_3
&=& \sum_{\omega\kv\qv} B_0 \frac{1}{2} f'(\omega) 
\left(
\ga_{\kv-\frac{q}{2},\mp,\omega}
\ga_{\kv+\frac{q}{2},\pm,\omega}
+ \gr_{\kv-\frac{q}{2},\mp,\omega}
\gr_{\kv+\frac{q}{2},\pm,\omega}
\right)
\nonumber\\
S_4 &=& 
- \sum_{\omega\kv\qv} \frac{1}{2} B_0 f(\omega) 
\nonumber\\ &&
\times \left[
 (\ga_{\kv-\frac{q}{2},\mp,\omega})^2 
 \ga_{\kv+\frac{q}{2},\pm,\omega}
- 
 \ga_{\kv-\frac{q}{2},\mp,\omega} 
 (\ga_{\kv+\frac{q}{2},\pm,\omega})^2
-(\gr_{\kv-\frac{q}{2},\mp,\omega})^2 
 \gr_{\kv+\frac{q}{2},\pm,\omega}
+
 \gr_{\kv-\frac{q}{2},\mp,\omega}
 (\gr_{\kv+\frac{q}{2},\pm,\omega})^2
\right]
\nonumber\\
S_5  &=&
\sum_{\omega\kv\qv} \frac{1}{2} B_0 f(\omega) 
\left[
\ga_{\kv-\frac{q}{2},\mp,\omega} \vvec{\partial}_{\omega} 
 \ga_{\kv+\frac{q}{2},\pm,\omega}
- \gr_{\kv-\frac{q}{2},\mp,\omega} \vvec{\partial}_{\omega}
 \gr_{\kv+\frac{q}{2},\pm,\omega}
\right] 
\nonumber\\
&&+\sum_{\omega\kv\qv} \frac{1}{2} B_i q_i f(\omega) 
\left[
(\ga_{\kv-\frac{q}{2},\mp,\omega})^2
 (\ga_{\kv+\frac{q}{2},\pm,\omega})^2
- (\gr_{\kv-\frac{q}{2},\mp,\omega})^2
 (\gr_{\kv+\frac{q}{2},\pm,\omega})^2
\right)],
\end{eqnarray}
where
we used partial integration with respect to $k_i$ and
the identity
\begin{equation}
m\sum_i \partial_{k_i}B_i =B_0.
\end{equation}

The correction to the current vertex, eq. (\ref{sV}), is calculated as
\begin{eqnarray}
\stil^{\pm}(\xv,t)^{\rm (B)} & = &
\sum_{\omega\kv\qv} \frac{B_0}{\Omz} 
\left[
g_{\kv-\frac{q}{2},\mp,\omega-\frac{\Omz}{2}}  g_{\kv+\frac{q}{2},\pm,\omega+\frac{\Omz}{2}}\right]^<
\nonumber\\
&=& -S_2-S_3-S_4
+\sum_{\omega\kv\qv} {B_0} f'(\omega)
 \gr_{\kv-\frac{q}{2},\mp,\omega} 
 \ga_{\kv+\frac{q}{2},\pm,\omega}.
\end{eqnarray}
Thus the current-driven part is given by
\begin{eqnarray}
\stil^{\pm (1)}(\xv,t) & = & 
  \stil^{\pm}(\xv,t)^{\rm (A)} +\stil^{\pm}(\xv,t)^{\rm (B)} 
\nonumber\\
&=& 
\sum_{\omega\kv\qv} {B_0} f'(\omega)
 \gr_{\kv-\frac{q}{2},\mp,\omega} 
 \ga_{\kv+\frac{q}{2},\pm,\omega}
+S_1+S_5 \nonumber\\
&=&
\sum_{\omega\kv\qv} \sum_i B_i f'(\omega) 
\nonumber\\ && \times 
\left[ \left(k+\frac{q}{2}\right)_i 
 \gr_{\kv-\frac{q}{2},\mp,\omega} 
 \gr_{\kv+\frac{q}{2},\pm,\omega} 
 \ga_{\kv+\frac{q}{2},\pm,\omega}
+\left( k-\frac{q}{2}\right)_i 
\gr_{\kv-\frac{q}{2},\mp,\omega}
\ga_{\kv-\frac{q}{2},\mp,\omega} \ga_{\kv+\frac{q}{2},\pm,\omega}
\right]
\nonumber\\
&& +
\sum_{\omega\kv\qv} {B_0} f'(\omega)
 \gr_{\kv-\frac{q}{2},\mp,\omega} 
 \ga_{\kv+\frac{q}{2},\pm,\omega}
\nonumber\\
&& 
 +\sum_{\omega\kv\qv} \frac{1}{2} B_0 f(\omega) 
\left[
\ga_{\kv-\frac{q}{2},\mp,\omega} \vvec{\partial}_{\omega} 
 \ga_{\kv+\frac{q}{2},\pm,\omega}
- \gr_{\kv-\frac{q}{2},\mp,\omega} \vvec{\partial}_{\omega}
 \gr_{\kv+\frac{q}{2},\pm,\omega}
\right] 
\nonumber\\
&&+\sum_{\omega\kv\qv} \frac{1}{2} B_i q_i f(\omega) 
\left[
(\ga_{\kv-\frac{q}{2},\mp,\omega})^2
 (\ga_{\kv+\frac{q}{2},\pm,\omega})^2
- (\gr_{\kv-\frac{q}{2},\mp,\omega})^2
 (\gr_{\kv+\frac{q}{2},\pm,\omega})^2
\right)].  \nonumber\\
\end{eqnarray}

\subsection{Dominant term}
We consider, in this paper, a clean limit, 
${(\eF \tau)}^{-1}\rightarrow 0$, where $\tau$ is the elastic lifetime due to impurities, which we assume to be spin-independent.
In this limit, spin density is dominated by $S_1$.
Thus
\begin{eqnarray}
\stil^{\pm (1)}(\xv,t)  &=&
\sum_{\omega\kv\qv} \sum_i B_i f'(\omega) 
\nonumber\\ && \times 
\left[ \left(k+\frac{q}{2}\right)_i 
 \gr_{\kv-\frac{q}{2},\mp,\omega} 
 \gr_{\kv+\frac{q}{2},\pm,\omega} 
 \ga_{\kv+\frac{q}{2},\pm,\omega}
+\left( k-\frac{q}{2}\right)_i 
\gr_{\kv-\frac{q}{2},\mp,\omega}
\ga_{\kv-\frac{q}{2},\mp,\omega} \ga_{\kv+\frac{q}{2},\pm,\omega}
\right] \nonumber\\
&&
\end{eqnarray}
Since $\kb T/\eF \ll 1$, we can replace
$f'(\omega)\simeq -\delta(\omega)$, and hence
\begin{eqnarray}
\stil^{\pm (1)}(\xv,t)  &=&
-\frac{1}{2\pi}\sum_{\kv\qv} \sum_i B_i 
\nonumber\\ && \times 
\left[ \left(k+\frac{q}{2}\right)_i 
 \gr_{\kv-\frac{q}{2},\mp} 
 \gr_{\kv+\frac{q}{2},\pm} 
 \ga_{\kv+\frac{q}{2},\pm}
+\left( k-\frac{q}{2}\right)_i 
\gr_{\kv-\frac{q}{2},\mp}
\ga_{\kv-\frac{q}{2},\mp} 
\ga_{\kv+\frac{q}{2},\pm}
\right],
\nonumber\\ &&
\end{eqnarray}
where $\gr_{\kv-\frac{q}{2},\mp}\equiv \gr_{\kv-\frac{q}{2},\mp,\omega=0}$.
Using the identities
\begin{eqnarray}
 \gr_{\kv,\sigma} \ga_{\kv,\sigma} &=&  
 i\tau (\gr_{\kv,\sigma} - \ga_{\kv,\sigma})
\\
 \gr_{\kv-\frac{q}{2},\mp} \ga_{\kv+\frac{q}{2},\pm} &=&  
-\frac{1}
{\epsilon_{\kv+\frac{q}{2}}-\epsilon_{\kv-\frac{q}{2}} \mp2\Delta +i\gamma} (\gr_{\kv-\frac{q}{2},\mp} - \ga_{\kv+\frac{q}{2},\pm})
\\
 \gr_{\kv-\frac{q}{2},\mp} \gr_{\kv+\frac{q}{2},\pm} &=&  
-\frac{1}
{\epsilon_{\kv+\frac{q}{2}}-\epsilon_{\kv-\frac{q}{2}} \mp2\Delta } (\gr_{\kv-\frac{q}{2},\mp} - \gr_{\kv+\frac{q}{2},\pm}),
\end{eqnarray}
where $\gamma\equiv \frac{1}{\tau}$,
we obtain, in the limit of $\tau\rightarrow\infty$, 
\begin{eqnarray}
\stil^{\pm (1)}(\xv,t)  &=&
-\frac{i}{2\pi}\sum_{\kv\qv} \sum_i  \frac{eE_i}{2 m V} e^{-i\qv\cdot\xv} \sum_{\mu}
 A_\mu^\pm (\qv,0) \tau
\nonumber\\ && \times 
\left[ 
 k_i  J_\mu(-2(k+\frac{q}{2})) 
 \frac{\gr_{\kv,\pm} -\ga_{\kv,\pm} }
{\epsilon_{\kv+q}-\epsilon_{\kv} \pm2\Delta} 
-  k_i J_\mu(2(\kv+\frac{q}{2}))
 \frac{\gr_{\kv,\mp} - \ga_{\kv,\mp}}
{\epsilon_{\kv+{q}}-\epsilon_{\kv} \mp2\Delta}
\right.\nonumber\\
&& \left.
-i\pi q_i \left(
\gr_{\kv,\mp} J_\mu(2(\kv+\frac{q}{2})) 
\delta({\epsilon_{\kv+q}-\epsilon_{\kv} \mp2\Delta})
-\ga_{\kv,\pm}  J_\mu(-2(\kv+\frac{q}{2})) 
\delta({\epsilon_{\kv+q}-\epsilon_{\kv} \pm2\Delta})
\right)
\right]
\nonumber\\ \label{S1_2}
\end{eqnarray}

\section{Estimation of $\stilpara$ and $\stilperp$}
We carry out the estimation of 
$\stilpara\equiv \half \sum_{\pm} e^{\mp i \phi}\stil^{\pm}$ and 
$\stilperp \equiv \half\sum_{\pm} \mp i e^{\mp i \phi}\stil^{\pm}$.
From eq. (\ref{S1_2}), we obtain
\begin{eqnarray}
\stilpara^{(1)}(\xv,t)&=&
 \half \sum_{\pm}\sum_{\kv\qv} \sum_{i}  
  \frac{-ieE_i\tau}{4\pi m V}
 e^{-i\qv\cdot\xv}
\nonumber\\ && \times 
\left[ 
-\frac{2}{m}k_i \left(k+\frac{q}{2}\right)_j
  (\bar{A}_j^\pm+\bar{A}_j^\mp)
 \frac{\gr_{\kv,\pm} -\ga_{\kv,\pm} }
{\epsilon_{\kv+q}-\epsilon_{\kv} \pm2\Delta} 
\right.\nonumber\\
&& \left.
-i\pi q_i \delta({\epsilon_{\kv+q}-\epsilon_{\kv} \pm2\Delta})
\frac{2}{m} \left(k+\frac{q}{2}\right)_j
\left(
\gr_{\kv,\pm}\bar{A}_j^\mp+ \ga_{\kv,\pm} \bar{A}_j^\pm \right) 
\right],  \nonumber\\
    \label{S1para}
\end{eqnarray}
where
$\bar{A}_\mu^\pm(\xv,\qv) \equiv e^{\mp i\phi(\xv)}  A_\mu^\pm (\qv)$
and
\begin{eqnarray}
\stilperp^{(1)}(\xv,t)
&=& 
 \half \sum_{\pm}\sum_{\kv\qv} \sum_{i\mu} (\mp)\frac{eE_i\tau}{4\pi mV}
 e^{-i\qv\cdot\xv}
\nonumber\\ && \times 
\left[ 
 -\frac{2}{m}k_i \left(k+\frac{q}{2}\right)_j (\bar{A}_j^\pm-\bar{A}_j^\mp) 
 \frac{\gr_{\kv,\pm} -\ga_{\kv,\pm} }
{\epsilon_{\kv+q}-\epsilon_{\kv} \pm2\Delta} 
\right.\nonumber\\
&& \left.
+i\pi q_i \delta({\epsilon_{\kv+q}-\epsilon_{\kv} \pm2\Delta})
\frac{2}{m} \left(k+\frac{q}{2}\right)_j
\left(
\gr_{\kv,\pm}\bar{A}_j^\mp - \ga_{\kv,\pm} \bar{A}_j^\pm \right) 
\right]  .    \nonumber\\
 \label{S1perp}
\end{eqnarray}
From the symmetry argument, we see that
$k_i \left(k+\frac{q}{2}\right)_j \rightarrow
\frac{\delta_{ij}}{3}
\kv\cdot\left(\kv+\frac{\qv}{2}\right)$
and
$q_i \left(k+\frac{q}{2}\right)_j \rightarrow
\frac{\delta_{ij}}{3}
\qv\cdot\left(\kv+\frac{\qv}{2}\right)$.
We thus obtain
\begin{eqnarray}
\stilpara^{(1)}(\xv,t)&=&
  \frac{e\tau}{6\pi m^2 V}
\sum_{\kv\qv\pm} e^{-i\qv\cdot\xv} i(\gr_{\kv,\pm}-\ga_{\kv,\pm} )
\left[ 
(\Ev\cdot \bar{\Av}^{\rm (s)}) 
\kv\cdot \left(\kv+\frac{\qv}{2}\right)
 \frac{1}
{\epsilon_{\kv+q}-\epsilon_{\kv} \pm2\Delta} 
\right.\nonumber\\
&& \left.
\pm \pi (\Ev\cdot \bar{\Av}^{\rm (a)}) \qv\cdot\left(\kv+\frac{\qv}{2}\right)
\delta({\epsilon_{\kv+q}-\epsilon_{\kv} \pm2\Delta})
\right]      \label{S1para1}
\nonumber\\
\stilperp^{(1)}(\xv,t)&=&
  \frac{e\tau}{6\pi m^2 V}
\sum_{\kv\qv\pm} e^{-i\qv\cdot\xv} i(\gr_{\kv,\pm}-\ga_{\kv,\pm} )
\left[ 
(\Ev\cdot \bar{\Av}^{\rm (a)}) 
\kv\cdot \left(\kv+\frac{\qv}{2}\right)
 \frac{1}
{\epsilon_{\kv+q}-\epsilon_{\kv} \pm2\Delta} 
\right.\nonumber\\
&& \left.
\mp \pi (\Ev\cdot \bar{\Av}^{\rm (s)}) \qv\cdot\left(\kv+\frac{\qv}{2}\right)
\delta({\epsilon_{\kv+q}-\epsilon_{\kv} \pm2\Delta})
\right] , \nonumber\\
\end{eqnarray}
where
\begin{eqnarray}
\bar{A}_\mu^{s}(\xv,\qv) 
  &\equiv& \half(\bar{A}_\mu^+ +\bar{A}_\mu^-)\nonumber\\
&=& \evsph(\xv)\cdot \Av_\mu(\qv) \nonumber\\
\bar{A}_\mu^{a} (\xv,\qv)
  &\equiv& -i\half(\bar{A}_\mu^+ -\bar{A}_\mu^-)\nonumber\\
&=& \evph(\xv)\cdot \Av_\mu(\qv) .
\end{eqnarray}
Thus we obtain eq. (\ref{stils}).


\end{document}